\documentclass[preprint,10pt]{elsarticle}

\usepackage{amssymb,amsmath}
\numberwithin{equation}{section}
\biboptions{sort&compress}

\def\be{\begin{equation}}
\def\ee{\end{equation}}

\usepackage{graphicx,eucal}
\usepackage[all]{xy}
\xyoption{all}

\usepackage{color}
\usepackage{subfigure}

\journal{Physica A}

\begin{document}

\begin{frontmatter}



\title{A First Look at First-Passage Processes}
\author{S. Redner}
\ead{redner@santafe.edu}
\address{Santa Fe Institute, 1399 Hyde Park Road, Santa Fe, New Mexico 87501,
  USA}

\begin{abstract}
  These notes are based on the lectures that I gave (virtually) at the
  Bruneck Summer School in 2021 on first-passage processes and some
  applications of the basic theory.  I begin by defining a
  first-passage process and presenting the connection between the
  first-passage probability and the familiar occupation probability.
  Some basic features of first passage on the semi-infinite line and a
  finite interval are then discussed, such as splitting probabilities
  and first-passage times.  I also treat the fundamental connection
  between first passage and electrostatics.  A number of applications
  of first-passage processes are then presented, including the hitting
  probability for a sphere in greater than two dimensions, reaction
  rate theory and its extension to receptors on a cell surface,
  first-passage inside an infinite absorbing wedge in two dimensions,
  stochastic hunting processes in one dimension, the survival of a
  diffusing particle in an expanding interval, and finally the
  dynamics of the classic birth-death process.

\end{abstract}

\end{frontmatter}

\section{What is a First-Passage Process?}

The \emph{first-passage probability} is defined as the probability that a
diffusing particle or a random walk \emph{first} reaches a given site (or set
of sites) at a specified time.  Typical examples of first-passage processes
include: fluorescence quenching, in which light emission by a fluorescent
molecule stops when it reacts with a quencher; integrate-and-fire neurons, in
which a neuron fires when a fluctuating voltage level first reaches a
specified level; and the execution of buy/sell orders when a stock price
first reaches a threshold.  To appreciate why first-passage phenomena might
be relevant practically, consider the following example.  You are an investor
who buys stock in a company at a price of \$100.  Suppose that this price
fluctuates daily by $\pm \$1$.  You will sell if the stock price reaches
\$200 and if the stock price reaches \$0, the company has gone bankrupt and
you've lost all your investment.  What it the probability of doubling your
investment or losing your entire investment?  How long will it take before
one of these two events occurs?  These are the types of questions that are
the purview of first-passage phenomena.  Much of the material covered here is
discussed more detail in this monograph~\cite{fpp}, and in other general
reviews and texts on probability theory and stochastic
processes~\cite{BMS15,F68,K97,KT14}.

\section{First-Passage and Occupation Probabilities}
\label{sec:c-fund-connect}

Let's start by deriving the formal relation between the first-passage
probability and the familiar occupation probability.  For concreteness,
consider a random walk in discrete space and discrete time.  We define
$P(\mathbf{r},t)$ as the occupation probability; this is the probability that
a random walk is at site $\mathbf{r}$ at time $t$ when it starts at the
origin.  Similarly, let $F(\mathbf{r},t)$ be the first-passage probability,
namely, the probability that the random walk \emph{first} visits $\mathbf{r}$
at time $t$ with the same initial condition.  Clearly $F(\mathbf{r},t)$
decays more rapidly in time than $P(\mathbf{r},t)$ because once a random walk
reaches $\mathbf{r}$, there can be no further contribution to
$F(\mathbf{r},t)$, although the same walk may still contribute to
$P(\mathbf{r},t)$.

\begin{figure}[ht]
\begin{center}
\includegraphics[width=0.5\textwidth]{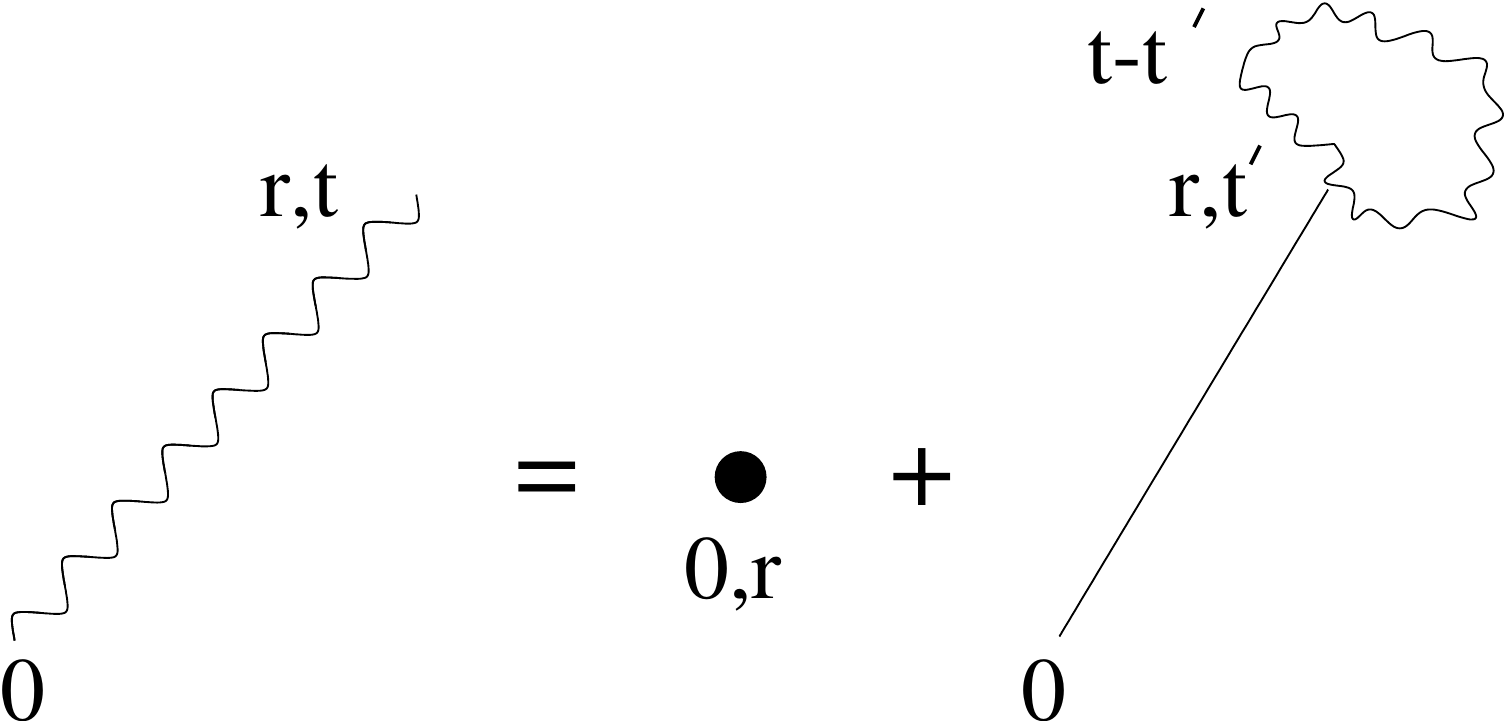}
\caption{Schematic diagrammatic relation between the occupation probability
  of a random walk (whose propagation is represented by a wavy line) and the
  first-passage probability (straight line).}~\label{fig-c-fund-convolution}
\end{center}
\end{figure}

It is convenient to write $P(\mathbf{r},t)$ in terms of $F(\mathbf{r},t)$ and
then invert this relation to find $F(\mathbf{r},t)$.  For a random walk to be
located at $\mathbf{r}$ at time $t$, the walk must first reach $\mathbf{r}$
at some earlier time step $t'$ and then return to $\mathbf{r}$ after $t-t'$
additional steps (Fig.~\ref{fig-c-fund-convolution}).  This connection
between $F(\mathbf{r},t)$ and $P(\mathbf{r},t)$ may be expressed as the
convolution
\begin{equation}
\label{c-fund:fund}
P(\mathbf{r},t)=\delta_{\mathbf{r},0}\delta_{t,0} + \sum_{t'\leq t} F(\mathbf{r},t')\,P(0,t-t')\,,
\end{equation}
where $\delta_{t,0}$ is the Kronecker delta function.  This equation
expresses the fact that if a random walk is at $\mathbf{r}$ at time $t$, it
must have first reached $\mathbf{r}$ at some earlier time $t'$ (which could
even be $t$).  If the walk reached $\mathbf{r}$ at a time earlier than $t$,
then it must return to $\mathbf{r}$ (and any number of such returns could
occur) in the remaining time $t-t'$.  The probability for this set of events
is expressed by $P(0,t-t')$.  The delta function term accounts for the
initial condition that the walk starts at $\mathbf{r}=0$. 

The above convolution is most conveniently solved by introducing the
generating functions,
\begin{align*}
  P(\mathbf{r},z)=\sum_{t=0}^\infty P(\mathbf{r},t)z^t\,,
  \qquad F(\mathbf{r},z)=\sum_{t=0}^\infty F(\mathbf{r},t)z^t\,.
\end{align*}
For a random walk in continuous time, we would merely replace the sum over
discrete time in Eq.~\eqref{c-fund:fund} by an integral and then use the
Laplace transform.  However, the asymptotic results would be identical.  To
solve for the first-passage probability, we multiply Eq.~(\ref{c-fund:fund})
by $z^t$ and sum over all $t$.  We thereby find that the generating functions
for $P$ and $F$ are related by
\begin{equation}
\label{c-fund:gfz}
P(\mathbf{r},z)=\delta_{\mathbf{r},0}+F(\mathbf{r},z)P(0,z)\,.
\end{equation}
Thus we obtain the fundamental connection between the generating functions
\begin{equation}
\label{c-fund:ret}
F(\mathbf{r},z)=
\begin{cases}
  {\displaystyle \frac{P(\mathbf{r},z)}{P(0,z)}}& \mathbf{r}\ne 0\,,\\[4mm]
      {\displaystyle 1-\frac{1}{P(0,z)}}, & \mathbf{r}= 0\,.
\end{cases}
\end{equation}

The important point is that the first-passage probability can be determined
solely from the occupation probability.  Many profound results about random
walks in infinite space can be obtained from the fundamental relation
\eqref{c-fund:ret} (see, e.g., \cite{M65,MW65,W94,H95}).  Our focus here will
be on random walks or diffusion in confined geometries that reflect important
physical constraints.

\section{The Half Line}
\label{sec:half-line}

Suppose that a diffusing particle starts at $x_0>0$ on the infinite half line
$[0,\infty]$ and is absorbed when it reaches the origin.  Does the particle
ever reach the origin?  If so, when does this particle \emph{first} reach the
origin?  To answer these questions, we have to solve the diffusion equation
for the concentration $c(x,t)$, subject to the initial condition
$c(x,t\!=\!0)=\delta(x-x_0)$, and the boundary condition
$c(x\!=\!0,t\!>\!0)=0$; the latter enforces the absorption of the particle
when it reaches the origin. 

A standard approach to solve this problem is to first take the Laplace
transform of the diffusion equation and then solve for the Green's
function of this transformed equation. Then one inverts the Laplace
transform of the Green's function to obtain the concentration $c(x,t)$
in the time domain.  The diffusive flux to the origin gives the
probability that the particle reaches the origin at time $t$.  Because
of the absorbing boundary condition, when the particle does reach the
origin, it is removed from the system.  Thus the diffusive flux
corresponds to the probability for the particle to reach the origin
for the \emph{first} time---namely, the first-passage probability to
the origin.

\begin{figure}[ht]
  \begin{center}
    \includegraphics[width=0.5\textwidth]{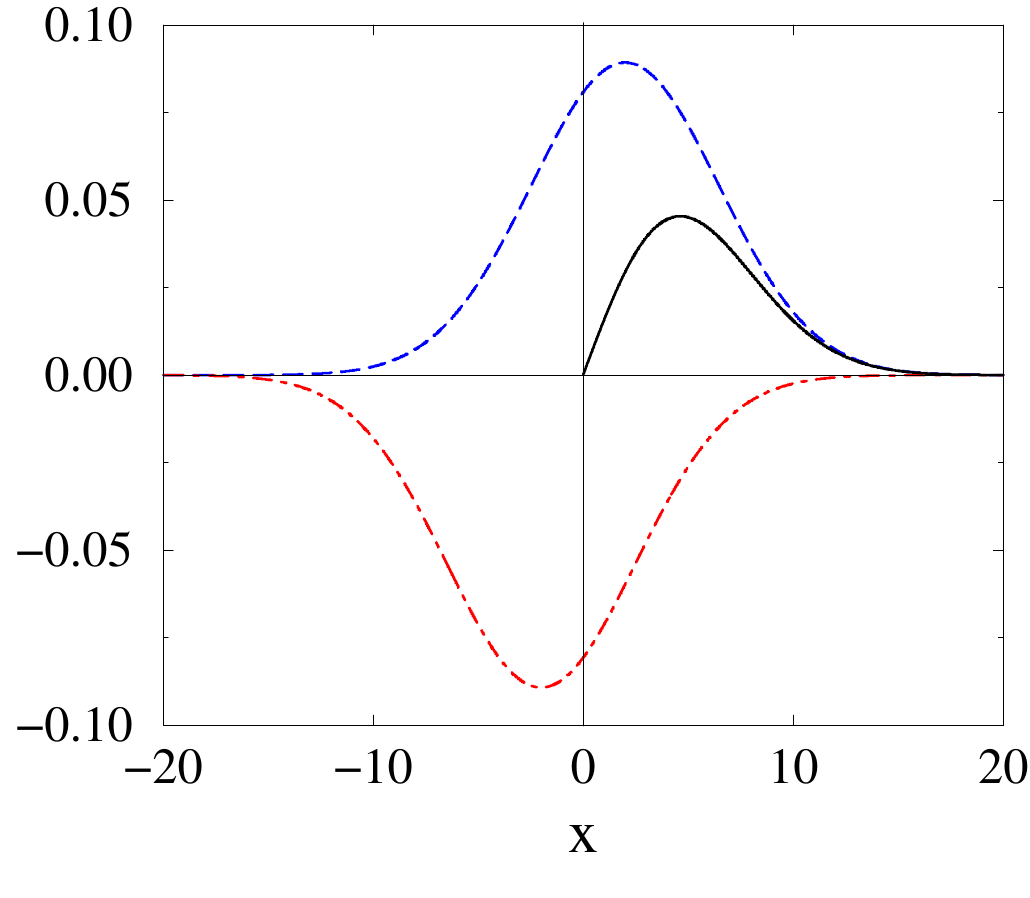}
    \caption{Concentration profile of a diffusing particle on the absorbing
      infinite half line $(0,\infty)$ at $Dt=10$, with $x_0=2$ (solid curve).
      Also shown are the component Gaussian (blue dashed) and image
      anti-Gaussian (red dot-dash) and their superposition in the physical
      region $x>0$.}~\label{fig:image}
  \end{center}
\end{figure}

A more fun way to solve this problem is by invoking the familiar image method
from electrostatics.  In this method, a diffusing particle that starts at
$x_0>0$ and is subject to an absorbing boundary condition at $x=0$ is
equivalent to removing the boundary altogether and introducing an image
``antiparticle'' that is initially at $x=-x_0$.  Because the concentration is
clearly equal to zero at the origin by symmetry, this image antiparticle
effectively imposes the absorbing boundary condition $c(x\!=\!0,t)=0$.  The
initial particle and the image antiparticle both diffuse freely on
$[-\infty,\infty]$ and their superposition gives a resultant concentration
$c(x,t)$ for $x>0$ that solves the original problem.

Hence the concentration for a diffusing particle on the positive half-line is
the sum of a Gaussian centered at $x_0$ and an anti-Gaussian centered at
$-x_0$:
\begin{equation}
  \label{c-semi-inf:conca}
c(x,t)=\frac{1}{\sqrt{4\pi Dt}}\left[e^{-(x-x_0)^2/4Dt}- e^{-(x+x_0)^2/4Dt}\right].
\end{equation}
This concentration profile has a linear dependence on $x$ near the origin and
a Gaussian tail for $x/\sqrt{Dt}\gg 1$, as illustrated in
Fig.~\ref{fig:image}.  Because the initial condition is normalized, the
first-passage probability to the origin at time $t$ is just the diffusive
flux to this point.  From the above expression for $c(x,t)$, we find
\begin{align}
  \label{c-semi-inf:fpp}
F(0,t) = +D\frac{\partial c(x,t)}{\partial x}\bigg|_{x=0} 
  =   \frac{x_0}{\sqrt{4\pi Dt^3}}\,\,e^{-x_0^2/4Dt}\qquad t\to\infty\,. 
\end{align}

This fundamental and simple formula has a number of striking implications:

\begin{enumerate}

\item The particle is sure to reach to the origin because
  $\int_0^\infty F(0,t)\, dt =1$.  That is, \emph{eventual absorption is
    certain}.

\item The \emph{average} time for the particle to reach the origin is
  infinite!  This fact arises because the first-passage probability has the
  long-time algebraic tail $F(0,t)\to x_0/\sqrt{4\pi Dt^3}$ for
  $t\gg x_0^2/D$.  This dichotomy between hitting the origin with certainty
  but taking an infinite average time to do so underlies many of the
  intriguing features of one-dimensional diffusion.

\item The \emph{typical} time to reach the origin is finite.  We can define
  the term typical time in a precise way as follows: As a preliminary, define
  the typical position of the particle, $x_T$, as
\begin{align}
  \int_{x_T}^\infty c(x,t)\, dx=\tfrac{1}{2}\,.
\end{align}
That is, one half of the total probability lies in the range beyond $x_T$ and
one half lies in the range $[0,x_T]$.  Substituting the concentration profile
\eqref{c-semi-inf:conca} into the above integral and performing the integral
leads to the transcendental equation

\begin{align}
  \text{erfc}\left(\frac{x_T-x_0}{\sqrt{4Dt}}\right)-
  \text{erfc}\left(\frac{x_T+x_0}{\sqrt{4Dt}}\right)= 1\,.
\end{align}
This equation can only be solved numerically and the salient feature is that
$x_T$ monotonically decreases with time and reaches zero at a time that is
roughly $1.1\, x_0^2/D$; this defines the typical hitting time.

\item Even though the average time to reach the origin is infinite, the
  number of times the origin is reached in a time $t$ is proportional to
  $\sqrt{t}$.  This result follows directly from the probability distribution
  of a freely diffusing particle.  At large times, the bulk of Gaussian
  distribution for a particle that starts at $x_0$ will spread past the
  origin.  Each site that is within the Gaussian envelope will have been
  visited of the order of $\sqrt{t}$ times.  This last fact will be relevant
  for our discussion of the reaction rate of the sphere in
  Sec.~\ref{sec:sphere}.

\end{enumerate}
  
\section{The Finite Interval}

We now turn to the first-passage properties in a finite interval.  The reason
for focusing on the finite interval is that the basic first-passage questions
in this geometry have many profound implications.  Moreover, the interval
geometry is sufficiently simple that many results can be readily derived.
Let us begin by outlining the basic questions that we will address.  We
consider a diffusing particle that starts at some point $x_0$ within the
interval $[0,L]$, with absorbing boundary conditions at both ends of the
interval.  Eventually the particle is absorbed, and our goal is to
characterize the time dependence of this absorption.  Basic first-passage
questions include:
\begin{enumerate}
  
\item What is the time dependence of the survival probability $S(t)$?  This
  is the probability that a diffusing particle does not touch either
  absorbing boundary before time $t$.
  
\item What is the time dependence of the first-passage, or exit
  probabilities, to either 0 or to $L$ as a function of $x_0$?  Integrating
  these probabilities over all time gives the eventual hitting, or
  \emph{splitting probability} to a specified boundary.  What is the
  dependence of the splitting probability to 0 or to $L$ as a function of the
  starting position?
  
\item What is the average exit time, that is, the average time until the
  particle hits either of the absorbing boundaries as a function of starting
  position?  What are the \emph{conditional} exit times, that is, the average time
  to hit a specified boundary (without ever touching the other boundary) as
  a function of the starting position?

\end{enumerate}

To answer the first question, we need to solve the diffusion equation in the
interval, subject to the initial condition that a particle starts at $x_0$,
and with absorbing boundary conditions at both ends.  This is a standard
exercise and the result for the concentration is
\begin{equation}
\label{cxt-series}
c(x,t)=\sum_{n\geq 1}^\infty \frac{2}{L} \sin \frac{n\pi x_0}{L}\,  \sin \frac{n\pi x}{L}\;e^{-n^2\pi^2\,Dt/L^2}~.
\end{equation}
Since the large-$n$ eigenmodes decay more rapidly in time, the most slowly
decaying eigenmode dominates in the long-time limit.  As a result, the
survival probability, which is the spatial integral of the concentration over
the interval, asymptotically decays as
\begin{equation}
 \label{c-interval-diff:Svst}
S(t) \sim e^{-D\pi^2 t/L^2}\,.
\end{equation}

For answering the second question, it is mathematically simpler to work in
the Laplace transform domain.  Applying the Laplace transform to the
diffusion equation recasts it as
\begin{equation}
\label{c-interval-diff:ode}
sc(x,s)-c(x,t=0)=Dc''(x,s)\,,
\end{equation}
where the prime denotes differentiation with respect to $x$.  The argument
$s$ indicates that $c(x,s)$ is the Laplace transform.  Within the standard
Green's function approach, the homogeneous equation in each subdomain
$(0,x_0)$ and $(x_0,L)$ has elementary solutions of the form
$c(x,s)=A\exp(x\sqrt{{s/D}})+B\exp(-x\sqrt{{s/D}})$, with the constants $A$
and $B$ determined by the boundary conditions.  Because the absorbing
boundary condition at $x=0$ mandates an antisymmetric combination of
exponentials and because the form of the Green's function as $x\to L$ must be
identical to that as $x\to 0$, we can be immediately write
\begin{align}
\label{c-interval-diff:cl}
  \begin{split}
c_<(x,s)&= A\sinh\left(\sqrt{\frac{s}{D}}\,x\right)   \hskip 0.5in x<x_0\,,  \\[2mm]
c_>(x,s)&= B\sinh\left(\sqrt{\frac{s}{D}}(L-x)\right) \quad  x>x_0\,,
\end{split}
\end{align}
for the subdomain Green's functions $c_<$ and $c_>$, where $A$ and $B$ are
constants.

We now impose the continuity condition $c_<(x_0,s)=c_>(x_0,s)$ and the jump
condition that is obtained by integrating Eq.~(\ref{c-interval-diff:ode}) over an
infinitesimal interval that includes $x_0$:
\[
c'(x,s)\big|_{x=x_0^+}-c'(x,s)\big|_{x=x_0^-}=-1/D\,,
\]
to finally obtain
\begin{equation}
\label{c-interval-diff:green-f}
c(x,s)= \frac{\displaystyle \sinh\left(\sqrt{\frac{s}{D}}\,x_<\right) 
  \sinh\left(\sqrt{\frac{s}{D}}(L-x_>)\right)}
{\displaystyle \sqrt{sD}\sinh\left(\sqrt{\frac{s}{D}}\,L\right)}\,.
\end{equation}
\index{Green's function for diffusion equation!finite interval}

From this Green's function, the Laplace transform of the fluxes at $x=0$ and
at $x=L$ are
\begin{subequations}
\begin{align}
\label{c-interval-diff:j-}
j_-(s|x_0) &\equiv +D\frac{\partial c(x,s)}{\partial x}\Bigg|_{x=0}
  = {  \displaystyle \sinh\left({\sqrt\frac{s}{D}}(L-x_0)\right)}\bigg/
{\displaystyle \sinh\left({\sqrt\frac{s}{D}}\,L\right)}\,.\\
\label{c-interval-diff:j+}
j_+(s|x_0) &\equiv -D\frac{\partial c(x,s)}{\partial x}\Bigg|_{x=L} =
 {\displaystyle  \sinh\left(\sqrt{\frac{s}{D}}\,x_0\right)}\bigg/
{\displaystyle \sinh  \left( \sqrt{\frac{s}{D}}\,L \right)}\,.
\end{align}
\end{subequations}
The subsidiary argument $x_0$ in $j$ emphasizes that the flux depends on the
initial particle position.  Since the initial condition is normalized, the
magnitude of the flux to each boundary is identical to the respective
first-passage probabilities.

For $s=0$, these Laplace transforms are just the time-integrated
first-passage probabilities to 0 and at $L$.  These quantities therefore
coincide with the respective splitting probabilities, ${\mathcal{E}}_-(x_0)$ and
${\mathcal{E}}_+(x_0)$, namely, the probabilities to eventually hit the left and the
right ends of the interval as a function of the initial
position $x_0$:
\begin{align}
\label{c-interval-diff:res}
\begin{split}
{\mathcal{E}}_-(x_0) &= j_-(s\!=\!0|x_0)=1-\frac{x_0}{L}\,,\\[2mm]
{\mathcal{E}}_+(x_0) &= j_+(s\!=\!0|x_0)=\frac{x_0}{L}\,.
\end{split}
\end{align}
Thus the splitting probabilities are given by an amazingly simple
formula---the probability of reaching one endpoint is just the fractional
distance to the other endpoint!

\begin{figure}[ht]
  \begin{center}
   \includegraphics[width=0.9\textwidth]{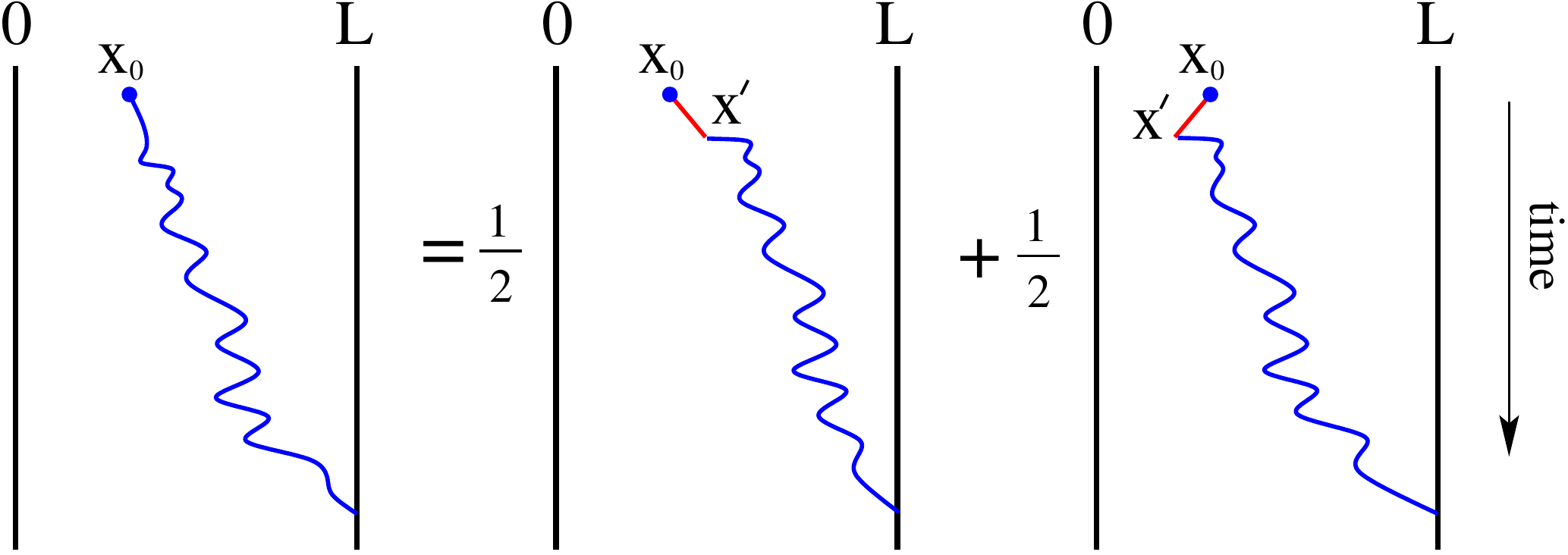}
   \caption{Schematic decomposition of a random walk path from $x_0$ to $L$
     into the outcome after one step (red) and the remainder from $x'$ to $L$.  The
     factors $1/2$ account for the probabilities associated with the first
     step of the decomposed paths.}~\label{fig:paths}
  \end{center}
\end{figure}

It is instructive to also derive these splitting probabilities by the
backward Kolmogorov approach~\cite{fpp,BMS15}.  The word backward reflects
the feature that the initial condition becomes the dependent variable, rather
than the current position of the particle.  As we shall see, this method
provides a powerful tool for determining first-passage properties.
Physically, we obtain the eventual hitting probability ${\mathcal{E}}_+(x_0)$ to the
right boundary by summing the probabilities for all paths that start at $x_0$
and reach $L$ without touching $0$.  Thus 
\begin{equation}
  \label{c-fund:poisson-exit-rec}
{\mathcal{E}}_+(x_0)=\sum_{\text{paths}} \Pi_{x_0\to L}\,,
\end{equation}
where $\Pi_{x_0\to L}$ denotes the probability of a path from $x_0$ to $L$ that
avoids $0$.  As illustrated in Fig.~\ref{fig:paths}, the sum over all such
paths can be decomposed into the outcome after one step and the sum over all
path remainders from the intermediate point $x'$ to $L$.  This gives
\begin{align}
  \label{c-fund:poisson-exit-rec-1}
\mathcal{E}_+(x_0) = \tfrac{1}{2}\sum_{\text{paths}'} 
\Pi_{x_0+\delta x\to L}+\tfrac{1}{2}\,\sum_{\text{paths}''} \Pi_{x_0-\delta x\to L}
  = \tfrac{1}{2}[{\mathcal{E}}_+(x_0\!+\!\delta x)+{\mathcal{E}}_+(x_0\!-\!\delta x)]\,.
\end{align}
Here $\delta x$ is the length of a single random-walk step and
paths$'$ and paths$''$ indicate, respectively, all paths that start at
$x_0\pm\delta x$ and reach $L$ without touching 0.

Equation~\eqref{c-fund:poisson-exit-rec-1}
reduces to $\Delta^{(2)} {\mathcal{E}}_\pm(x_0) =0$, where
$\Delta^{(2)}$ is the discrete second-difference operator,
$\Delta^{(2)} f(x)\equiv f(x-\delta x)-2f(x)+f(x+\delta x)$. This
difference equation is subject to the boundary conditions
${\mathcal{E}}_+(0)=0$, ${\mathcal{E}}_+(L)=1$.  The solution is
simply
\begin{align}
  \label{CE+}
   {\mathcal{E}}_+(x_0) =\frac{x}{L}\,,
\end{align}
and correspondingly ${\mathcal{E}}_-(x) = 1-\frac{x}{L}$.

It is worth mentioning that there is an even simpler way to determine
the exit probabilities by the martingale method~\cite{DS84}. This
solution relies on the fact that the motion of the random walk in the
interval is a ``fair game'' at any time.  Thus the average position of
the walk is time independent. Formally, a martingale is a process in
which the average value of a random variable at time $t+\delta t$
equals the average value of this variable at time $t$.

For the present example, at $t=0$, the average position of the
particle is $\langle x\rangle =x_0$.  At infinite time, the walk is
either at the left end or the right end of the interval, with
respective probabilities ${\mathcal{E}}_+(x_0)$ or
${\mathcal{E}}_-(x_0)$.  Thus the average position at infinite time is
$\langle x\rangle = 0 \times {\mathcal{E}}_-(x_0)+ L\times
{\mathcal{E}}_+(x_0)$.  Since the initial average position equals the
final average position, we immediately recover \eqref{CE+}.

\begin{figure}[ht]
  \begin{center}
    \includegraphics[width=0.6\textwidth]{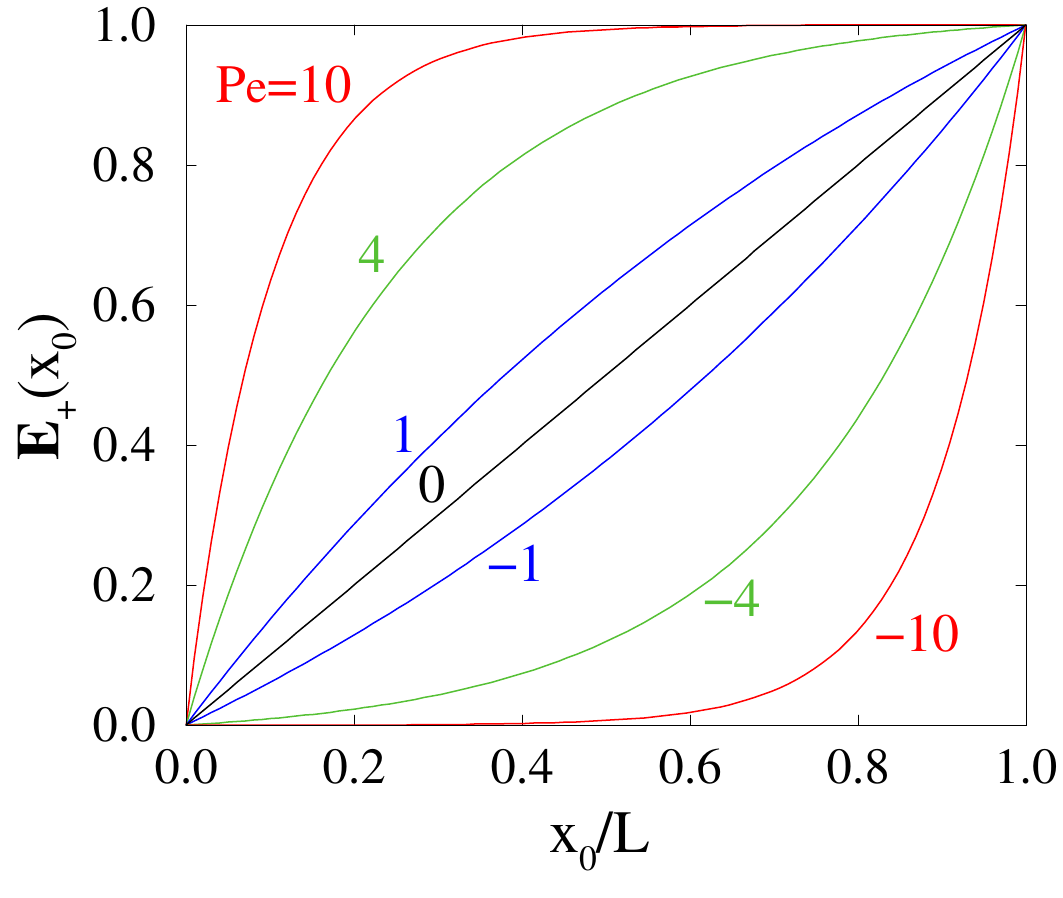}
    \caption{Dependence of the exit probability to $x=L$ on $u_0=x_0/L$ in
      the interval $[0,L]$ for various P\'eclet numbers
      $P\!e$.}~\label{fig:exit}
  \end{center}
\end{figure}

For a biased random walk with a probability $p$ of hopping to the right and
probability $q$ of hopping to the left, the analog of
Eq.~\eqref{c-fund:poisson-exit-rec-1} for the splitting probability is
\begin{align}
  \label{CE-bias}
{\mathcal{E}}(x_0) = p{\mathcal{E}}_+(x_0\!+\!\delta x)+q{\mathcal{E}}_+(x_0\!-\!\delta x)\,,
\end{align}
with solution
\begin{align}
  {\mathcal{E}}_+(x_0) = \frac{1-e^{-vx_0/D}}{1-e^{-vL/D}}
  \equiv  \frac{1-e^{-u_0P\!e}}{1-e^{-P\!e}}\,,
\end{align}
where $v=(p-q)\delta x$, $D=\delta x^2/2$, $u_0=x_0/L$, and $P\!e= vL/D$ is
the \emph{P\'eclet} number, which is a dimensionless measure of the influence
of the bias relative to diffusive fluctuations.  When the P\'eclet number is
large, the exit probability clearly reflects the strong influence of the bias
(Fig.~\ref{fig:exit}).

We now extend the backward Kolmogorov approach to determine the average exit
time from the finite interval.  We distinguish between the {\em
  unconditional\/} average exit time, namely, the average time for a particle
to reach \emph{either} end of the interval, and the {\em conditional\/}
average exit time, namely, the average time for a particle to reach, say, the
right end of the interval {\em without\/} ever touching the other end.

\begin{figure}[ht]
  \begin{center}
\includegraphics[width=0.6\textwidth]{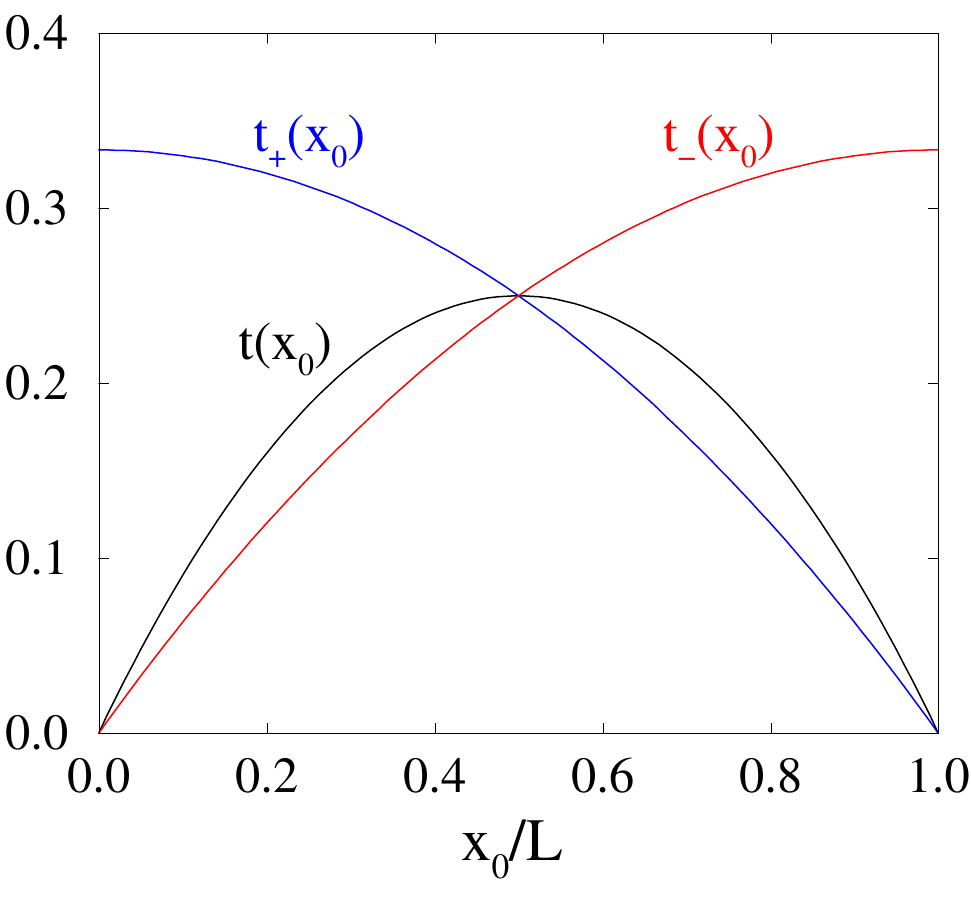}
 \caption{Unconditional and conditional average exit times from the finite interval
   $[0,L]$, normalized by $L^2/2D$, as a function of the dimensionless
   initial position $x_0/L$.}~\label{fig:times}
  \end{center}
\end{figure}

In close analogy with \eqref{c-fund:poisson-exit-rec}, the
unconditional exit time satisfies
\begin{equation}
  \label{t-un}
t(x_0)=\sum_{\text{paths}} (\Pi t)_{x_0\to \pm L}\,.
\end{equation}
That is, to compute the average unconditional exit time, we take the time for
a path to go from $x_0$ to $\pm L$ times the probability of this path and
sum over all possible paths.  Using the same decomposition that which led to
\eqref{c-fund:poisson-exit-rec-1}, the unconditional exit time satisfies
\begin{align}
  \label{t-un2}
t(x_0) &= \tfrac{1}{2}\!\!\sum_{\text{paths}'} \!\!
  \Pi_{x_0+\delta x\to \pm L}(t_{x_0+dx\to \pm L}\!+\!dt)
         +\tfrac{1}{2}\!\!\sum_{\text{paths}''}\!\! \Pi_{x_0-\delta x\to \pm L}
         (t_{x_0-dx\to \pm L}\!+\!dt)\,.
\end{align}
Notice that the term
\begin{align*} 
 \tfrac{1}{2}\!\!\sum_{\text{paths}'} \!\!
  \Pi_{x_0+\delta x\to \pm L}\; t_{x_0+dx\to \pm L}
\end{align*}
has exactly the same form as \eqref{t-un}, so that the above
expression is merely $\frac{1}{2} t(x_0+\delta x)$.  A similar
identification holds for the analogous term
\begin{align*} 
 \tfrac{1}{2}\!\!\sum_{\text{paths}''} \!\!
  \Pi_{x_0-\delta x\to \pm L}\; t_{x_0-dx\to \pm L}\,.
\end{align*}
Finally, the terms multiplying the factor $dt$ in \eqref{t-un2} just
gives the probability of \emph{all} possible paths from $x_0$ to
either end of the interval; this probability is clearly equal to 1.  Thus we have
\begin{align}
\label{t-un-final}
t(x_0)= \tfrac{1}{2}[t(x_0\!+\!\delta x)+t(x_0\!-\!\delta x)]+dt\,.
\end{align}
In the continuum limit, we expand \eqref{t-un-final} in a Taylor
series to second order in $\delta x$ to give
\begin{align*}
  \frac{\delta x^2}{2} t'' = -dt\,.
\end{align*}
Now identifying $\delta x^2/(2 dt)$ as the diffusion coefficient $D$,
the equation for the unconditional exit time reduces to $D\,t'' =-1$,
subject to the boundary conditions $t(0)=t(L)=0$.  The solution is
(see Fig.~\ref{fig:times})
\begin{equation}
  t(x_0) = \frac{x_0(L-x_0)}{2D}\,.
\end{equation}
Notice that this exit time is of the order of $L$ for a particle that starts
a distance of the order of one from an absorbing boundary and of the order of
$L^2$ for a particle that starts near the middle of the interval.

Now let's calculate the \emph{conditional} exit time to the right boundary
when starting from $x_0$, $t_+(x_0)$.  By definition this conditional time is
\begin{align}
  t_+(x_0) = \frac{\sum_{\text{paths}_+} (\Pi t)_{x_0\to L}}{\sum_{\text{paths}_+}
  \Pi_{x_0\to L}}= \frac{\sum_{\text{paths}_+} (\Pi t)_{x_0\to L}}{{\mathcal{E}}_+(x_0)}\,.
\end{align}
That is, the conditional exit time is the time for a path to start at
$x_0$ and reach $L$ without touching 0 multiplies by the probability
of this path, summed over all such allowable paths.  Here the
subscript + on the word paths indicates that only paths that go from
$x_0$ to $L$ without touching 0 are included.  Since the total
probability of all these restricted paths is less than 1, we need to
divide by this total probability, ${\mathcal{E}}_+(x_0)$, to obtain
the properly normalized conditional exit time.  Thus, for the quantity
${\mathcal{E}}_+(x_0) t_+(x_0)$, we have
\begin{align}
{\mathcal{E}}_+(x_0) t_+(x_0) &= \sum_{\text{paths}_+} (\Pi t)_{x_0\to L}\nonumber\\
  &= \tfrac{1}{2}\!\! \sum_{\text{paths}'_{+}} \Pi(t_{x_0+dx\to L}+dt)
+ \tfrac{1}{2} \!\!  \sum_{\text{paths}''_{+}} \Pi(t_{x_0-dx\to L}+dt)\nonumber\\
  &= \tfrac{1}{2}\left[{\mathcal{E}}_+(x_0\!+\!dx)t_+(x_0\!+\!dx)\right]
    +\tfrac{1}{2}\left[{\mathcal{E}}_+(x_0\!-\!dx)t_+(x_0\!-\!dx)\right]\nonumber\\[1mm]
  & \hskip 2.2 in +{\mathcal{E}}_+(x_0)\,dt\,.
\end{align}
In the continuum limit,
the above equation reduces to $D({\mathcal{E}}_+ t_+)'' = -{\mathcal{E}}_+$, with solution
\begin{align}
  t_+(x_0) =\frac{L^2-x_0^2}{6D}\,.
\end{align}
From this result, we also immediately find $t_-(x_0) =t_+(L-x_0)$.  The
dependences of all the exit times on the starting position are illustrated in
Fig.~\ref{fig:times}.

\section{Connection with Electrostatics}

One of the alluring features of first-passage processes is its intimate
connection to electrostatics.  By this connection, one can recast an
electrostatic problem in a given geometry as a first-passage problem in the
same geometry.  With this perspective, it is possible to solve seemingly
difficult first-passage problems in a simple way by this electrostatic
connection.

To illustrate the basic principle, consider the following general problem.
Suppose that a diffusing particle starts at some point $\mathbf{r}_0$ inside
an arbitrary bounded domain.  At the boundary of this domain, the particle is
absorbed.  Eventually, all of the initial probability is absorbed on the
boundary and we ask: what is the exit probability at some arbitrary point
$\mathbf{r}_B$ on the domain boundary?  Formally, we have to solve the
diffusion equation in this domain, subject to the appropriate initial and
boundary conditions:
\begin{align}
  \label{prob}
\begin{cases}
\displaystyle{\frac{\partial c(\mathbf{r},t)}{\partial t}} = D\nabla^2c(\mathbf{r},t)\\[2mm]
c(\mathbf{r},t=0)=\delta(\mathbf{r}-\mathbf{r}_0)\\[1mm]
  c(\mathbf{r},t)\big|_{\mathbf{r}_B}=0
\end{cases}
\end{align}
Then the exit probability at the boundary point $\mathbf{r}_B$ is given by
\begin{align}
\label{ErB}
\mathcal{E}(\mathbf{r}_B) = - D \int_0^\infty
\frac{\partial c(\mathbf{r},t)}{\partial \hat{\mathbf{n}}}\bigg|_{\mathbf{r}_B} \, dt\,,
\end{align}
where $\hat{\mathbf{n}}$ is the outward normal to the surface of the domain
at $\mathbf{r}_B$.

Let's look critically at this calculation.  We are attempting to solve a
partial differential equation in some domain (which may well be difficult),
then take the result of this calculation and integrate over all time.  That
is, we really don't need that exit probability at all times, but merely the
time integral of the exit probability.  This observation suggests that it
will be useful to take the original problem \eqref{prob} and integrate it over
all time.  To simplify what emerges, we also define the time integrated
concentration, $\mathcal{C}(\mathbf{r})\equiv\int_0^\infty c(\mathbf{r},t)\, dt$.
Performing this time integration on Eq.~\eqref{prob} leads to
\begin{align}
  \begin{cases}
    -\delta(\mathbf{r}-\mathbf{r}_0)= D\nabla^2 \mathcal{C}(\mathbf{r})\nonumber\\
  \mathcal{C}(\mathbf{r})\big|_{\mathbf{r}_B} =0
\end{cases}
\end{align}
The delta function on the left-hand side is what remains when we integrate
the time derivative in \eqref{prob} over all time.  At $t=\infty$ the
concentration is zero, while at $t=0$, we merely have the initial condition.
But notice that in terms of the time-integrated concentration, the exit
probability may be written as
\begin{align}
  \label{ErB2}
  \mathcal{E}(\mathbf{r}_B) = - D \frac{\partial \mathcal{C}(\mathbf{r},t)}{\partial \hat{\mathbf{n}}}\bigg|_{\mathbf{r}_B}\,.
\end{align}
Thus we arrive at the fundamental result:
\begin{align*}
  \mathcal{E}(\mathbf{r}_B) &= \text{the\
electric\ field\ at\ } \mathbf{r}_B \text{\ on the surface of a grounded
conductor,}\nonumber\\ & \hskip 0.2in \text{when a point charge of magnitude\ } 1/(D\Omega_d) \text{\ is
  placed at\ }
  \mathbf{r}_0\,.
\end{align*}
Here $\Omega_d$ is the surface area of a $d$-dimensional sphere; this factor
is needed to convert the prefactor $D$ in the diffusion equation to the
correct prefactor in the Laplace equation.  Thus a given first-passage
problem can be expressed as an equivalent electrostatic problem in the same
geometry.

\section{Hitting a Sphere and Reaction Rate Theory}
\label{sec:sphere}

What is the probability $H(r)$ that a diffusing particle eventually hits a
sphere of radius $a$, when the particle starts at a distance $r>a$ from the
origin?  One can determine this hitting probability in the standard way by
solving the diffusion equation exterior to the sphere, computing the flux to
the sphere, and then integrating over all time.  However, it is much simpler
to use the connection between first passage and electrostatics.  Indeed, by a
direct extension of Eq.~\eqref{c-fund:poisson-exit-rec-1} to three
dimensions, the hitting probability satisfies (here for a discrete random
walk in Cartesian coordinates for simplicity)
\begin{align}
  H(x,y)&=\tfrac{1}{6}\big[ H(x+1,y,z)+H(x-1,y,z)+H(x,y+1,z)\nonumber\\
        & \hskip 0.5in + H(x,y-1,z)+H(x,y,z+1)+H(x,y,z-1)\big]\,.
\end{align}
Let us now take the continuum limit so that we can work in spherical
coordinates.  Then the above discrete difference equation becomes
$\nabla^2H=0$, subject to the boundary conditions $H(r\!=\!a)=1$ and
$H(\infty)=0$.  The solution is 
\begin{align}
  \label{H}
  H(r)= \frac{a}{r}\,.
\end{align}
Amazingly simple!

Now let's treat a related problem that is fundamental in chemical kinetics.
Suppose that there is an initially uniform concentration of particles
exterior to an absorbing sphere of radius $a$.  A fundamental kinetic
characteristic of the absorbing sphere is its \emph{reaction rate} $k$,
namely, the efficiency at which this sphere captures particles.  Formally,
$k$ is defined as the number of particles absorbed per unit time divided by
the initial concentration.  This normalization ensures that the reaction rate
is a quantity that is intrinsic to the system.  By dimensional analysis, the
reaction rate as defined above has units of (length)$^d/$time.  Moreover the
reaction rate can only be a function of intrinsic parameters of the system,
namely, the diffusion coefficient $D$ and the sphere radius $a$.  Since the
units of $D$ are length$^2$/time, we infer that $k$ must have units of
$D\,a^{d-2}$.  Thus the reaction rate $k=A\, D\, a^{d-2}$, where $A$ is a
constant of the order of 1.  This example shows the power of dimensional
analysis in obtaining a non-trivial physical property of multi-particle
system.

This simple result has some surprising implications.  First, in three
dimensions, the reaction rate is \emph{linear} in $a$; it is \emph{not}
proportional to the cross-sectional area of the sphere.  Second for $d<2$,
the reaction rate increases when the radius of the absorbing sphere is
decreased!  This nonsensical result indicates that there is a basic problem
with classic chemical kinetics for $d<2$.  This pathology arises because the
concentration field exterior to the absorber never reaches a steady state for
$d<2$.  Instead, the absorption rate of the sphere is time dependent.  In
contrast, for $d>2$, a steady state concentration field does arise.  Once
this steady state is reached, it is a trivial exercise to compute the
steady-state density by solving the Laplace equation rather than the
diffusion equation, and thereby obtain the steady-state flux to the sphere.
From these steps, one finds the exact reaction rate
\begin{align}
  k=4\pi Da\,.
\end{align}

\begin{figure}[ht]
  \begin{center}
\includegraphics[width=0.75\textwidth]{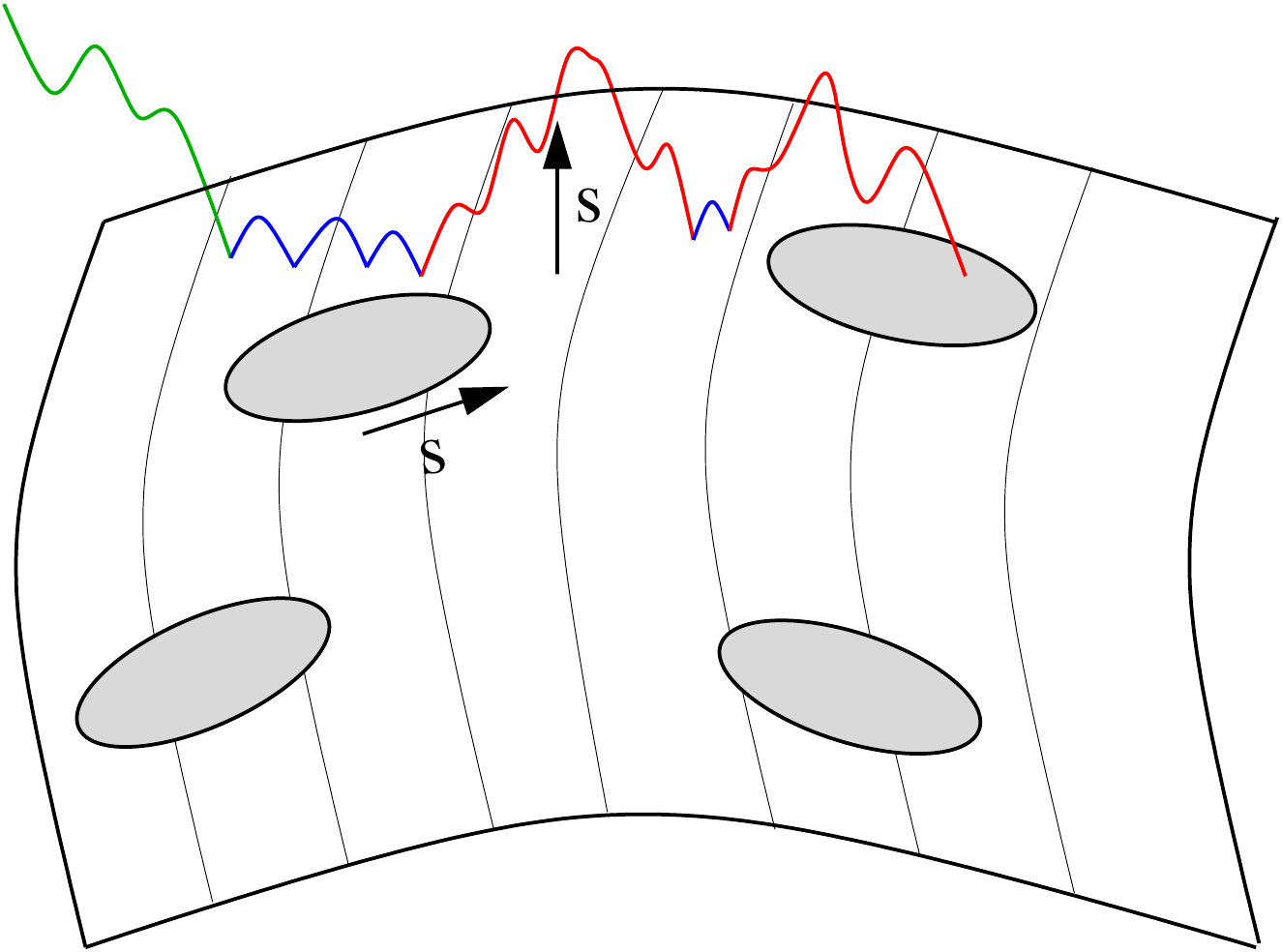}
\caption{Cartoon of a cell surface with 4 receptors (gray ovals).  A
  diffusing particle first hits the cell (green trajectory) and then makes 3
  non-independent hits of the surface (blue), before an independent hit
  (red).  The particle lands on a receptor after one more non-independent
  hit.}~\label{fig:hitting}
  \end{center}
\end{figure}

Armed with the basic result for the reaction rate, we now turn to a much more
profound problem that is fundamental to living systems, namely, how many
receptors should there be on the surface of a cell?  One might expect that
much of the cell surface should be covered by receptors so that its detection
efficiency is high.  On the other hand, one can imagine that receptors are
complex and evolutionarily expensive machines.  Based on cost considerations
only, it would be advantageous for a cell to minimize the number of
receptors.  What is the appropriate balance between these two competing
attributes?  As a first step to address this question, we want to compute the
reaction rate of a cell that is sensitive to its environment only at the
locations of the receptors.  We model the cell as a sphere of radius $a$ in
which most of the surface is reflecting.  However, on the sphere
surface there are also $N$ circular domains of radius $s$ that are absorbing
(Fig.~\ref{fig:hitting}).  We view these $N$ absorbing circles as the
receptors on the cell surface.  What is the reaction rate of this toy model
of a cell?  If most of the sphere surface is reflecting, one might anticipate
that the reaction efficiency of the cell will be poor.  Surprisingly, the
reaction efficiency of the sphere with $N$ absorbing receptors is almost as
good as a perfectly absorbing sphere, even when the area fraction covered by
the receptors is vanishingly small!  This realization is the brilliant
insight of the article by Berg and Purcell that was far ahead of its
time~\cite{BP77}.  Here I outline their argument.

The first step in their argument relies on the feature that if a diffusing
particle hits the surface of a sphere, it will hit again many times before
diffusing away; this point was discussed above in Sec.~\ref{sec:half-line}.
There are two types of subsequent hitting events: (i) the particle rises a
distance less than $s$ above the surface before hitting it again, and (ii)
the particle rises a distance greater than $s$ above the surface before
hitting it again (Fig.~\ref{fig:hitting}).  In the former case, if the
particle initially misses a receptor, it will likely miss upon the second
encounter.  In the latter case, if the particle misses a receptor initially,
we have no information about whether the second encounter will hit or miss a
receptor.  Thus the rise distance $s$ demarcates the regime of
\emph{dependent} subsequent hits and \emph{independent} subsequent hits.  It
is the latter events that are relevant for estimating the reaction rate.
We thus use the height $s$ as the criterion for determining the number of
times that a particle independently hits the cell surface before diffusing
away.

When a particle is a distance $s$ above the cell surface, the probability
that it eventually hits the cell again is, using Eq.~\eqref{H},
\begin{subequations}
\begin{align}
  p_s = \frac{a}{a+s}\,.
\end{align}
Thus the probability that the diffusing particle independently hits the cell
$n$ times before diffusing away is $p_s^n(1-p_s)$.  Correspondingly, the
average number of independent hits to the surface is
\begin{align}
  \langle n\rangle = \sum_{n>0} n p_s^n\;(1-p_s) = (1-p_s)^{-1} =
  \frac{a+s}{s}\sim \frac{a}{s}\,.
\end{align}

The probability for a diffusing particle to not land on a receptor in a
single independent hitting event is
\begin{align}
  \beta = 1 -\frac{N\pi s^2}{4\pi a^2}\,,
\end{align}
namely, the area fraction of the surface that is not covered by receptors.
Thus the probability $p_{\rm escape}$ that a diffusing particle that reaches
the cell but never lands on a receptor is the probability that the particle
always misses a receptor in each of its independent hitting attempts.  This
is
\begin{align}
  p_{\rm escape} = \sum_{n>0} (\beta p_s)^n(1-p_s) = \frac{1-p_s}{1-\beta
  p_s} = \frac{4a}{4a+Ns}\,.
\end{align}
The probability that a diffusing particle ultimately hits a receptor thus is
\begin{align}
  p_{\rm hit} = 1-p_{\rm escape} = \frac{Ns}{4a+Ns}\,.
\end{align}
\end{subequations}

The final result for the reaction rate of a cell of radius $a$ that is
covered by $N$ receptors, each of radius $s$, is
\begin{align}
  k= 4\pi Da \times \frac{Ns}{4a+Ns}\equiv 4\pi Da \times \text{efficiency}\,.
\end{align}
To understand the implication of this result, let's use some numbers that
typify a cell: $a= 1$ micron, $s=10$ nanometers, and $N=10^4$.  The area
fraction covered by the receptors is roughly $10^{-3}$, but the absorption
efficiency of the cell is roughly 2/3!  Evidently, Mother Nature is very
smart to not waste resources on endowing a cell with too many receptors.

\section{Wedge Domains}
\label{chap:wedge}

We now turn to the first-passage properties of diffusion in a two-dimensional
wedge domain with absorption when the particle hits the wedge boundary.  One
of our motivations for studying this system is that first passage in the
wedge can be mapped onto a simple diffusive capture problem in one dimension
that we'll treat in the next section.  We will obtain first-passage
properties in the wedge geometry both by direct solution of the diffusion
equation and also, for two dimensions, in a more aesthetically pleasing
fashion by conformal transformation techniques, in conjunction with the
electrostatic formulation.  We also present a heuristic extension of the
electrostatic approach that allows us to infer, with little additional
computational effort, time-dependent first-passage properties in the wedge
from corresponding time-integrated properties.

\subsection*{Solution to the Diffusion Equation} 

We first solve the diffusion equation in the wedge to determine the survival
probability of a diffusing particle.  While the exact Green's function for
this system is well known~\cite{S94,CJ59}, we adopt the strategy of choosing
an initial condition that allows us to eliminate angular variables and deal
with an effective radial problem.  This simplification is appropriate if we
are interested only in asymptotic first-passage properties.

The diffusion equation for the two-dimensional wedge geometry in plane polar
co-ordinates is
\begin{equation}
\label{c-domains:wedge-de}
\frac{\partial c}{\partial t} = D\left(\frac{\partial^2 c}{\partial r^2} + \frac{1}{r}
  \frac{\partial c}{\partial r} + \frac{1}{r^2}\frac{\partial^2 c}{\partial \theta^2}\right),
\end{equation}
where $c=c(r,\theta,t)$ is the particle concentration at $(r,\theta)$ at time
$t$, $D$ is the diffusion coefficient, and the boundary conditions are $c=0$
at $\theta=0,\Theta$, where $\Theta$ is the wedge opening angle.  To reduce
this two-dimensional problem to an effective one-dimensional radial problem,
note that the exact Green's function can be written as an eigenfunction
expansion in which the angular dependence is a sum of sine waves of the form
$\sin(n\pi\theta/\Theta)$, such that an integral number of half-wavelengths
fit within $(0,\Theta)$ to satisfy the absorbing boundary
conditions~\cite{CJ59}.  In this series, each sine wave is multiplied by a
conjugate decaying function of time, in which the decay rate increases with
$n$.  In the long time limit, only the lowest term in this expansion
dominates the survival probability.  Consequently, we obtain the long-time
behavior by choosing an initial condition whose angular dependence is a half
sine-wave in the wedge.  This ensures that the time-dependent problem will
contain only this single term in the Fourier series.

We therefore define $c(r,\theta,t=0)= \pi \delta(r-r_0)/(2\Theta r_0)\times
\sin(\pi\theta/\Theta)$.  With this initial distribution function and after the
Laplace transform is applied, the diffusion equation (\ref{c-domains:wedge-de})
becomes
\begin{align*}
sc(r,\theta,s)-\frac{\pi}{2\Theta r_0}\delta(r-r_0)\sin(\pi\theta/\Theta)= 
  D\left(\frac{\partial c^2}{\partial r^2}+\frac{1}{r}\frac{\partial c}{\partial r}+\frac{1}{r^2}
  \frac{\partial^2 c}{\partial \theta^2}\right)\,,
\end{align*}
where $c=c(r,\theta,s)$.  Substituting in the ansatz
$c(r,\theta,s)= R(r,s)\sin(\pi\theta/\Theta)$, the angular dependence may now
be separated and reduces the system to an effective one-dimensional radial
problem.  By introducing the dimensionless co-ordinate $x=r\sqrt{s/D}$, we
find the modified Bessel equation for the remaining radial co-ordinate,
\begin{equation}
\label{c-domains:wedge-de-r}
R''(x,s)+\frac{1}{x}R'(x,s) -\left(1+\frac{\nu^2}{x^2}\right)R(x,s)=
-\frac{\nu}{2Dx_0}\delta(x-x_0),
\end{equation}
where $\nu=\pi/\Theta$ and the prime now denotes differentiation with respect
to $x$.

The general solution for $x\ne x_0$ is a superposition of modified Bessel
functions of order $\nu$.  Since the domain is unbounded, the interior
Green's function ($x<x_0$) involves only $I_\nu$, since $K_\nu$ diverges as
$x\to 0$, while the exterior Green's function ($x>x_0$) involves only
$K_\nu$, since $I_\nu$ diverges as $x\to\infty$.  By imposing continuity at
$x=x_0$, we find that the Green's function has the symmetric form
$R(x,s)=AI_\nu(x_<)K_\nu(x_>)$, with the constant $A$ determined by the
joining condition that arises from integrating
Eq.~(\ref{c-domains:wedge-de-r}) over an infinitesimal radial range that
includes $r_0$.  This gives
\begin{align*}
R_>'\mid_{x=x_0}-R_<'\mid_{x=x_0}= -\frac{\nu}{2 Dx_0}\,,
\end{align*}
from which $A=\nu/2D$.  Therefore the radial Green's function in the wedge is
\begin{equation}
\label{c-domains:wedge-DF-soln}
R(x,s)=\frac{\nu}{2D}\, I_\nu(x_<)\, K_\nu(x_>),
\end{equation}
and its Laplace inverse has the relatively simple closed form~\cite{S94,CJ59}
\begin{equation}
\label{c-domains:wedge-GF-t}
R(r,t)= \frac{\nu}{4Dt}\,\,e^{-(r^2+r_0^2)/4Dt}\,\,I_\nu\left(\frac{rr_0}{2Dt}\right).
\end{equation}

With this radial Green's function, the asymptotic survival probability is
\begin{equation}
\label{c-domains:wedge-S-formal}
S(t)\sim\int_0^{\Theta} \sin(\nu\theta)\, d\theta\,
\int_0^\infty r\, R(r,t)\, dr.
\end{equation}
We can estimate this integral by noting that the radial distance over which
the concentration is appreciable extends to the order of $\sqrt{Dt}$.  This
provides a cutoff $r\approx \sqrt{Dt}$ in the radial integral in
Eq.~(\ref{c-domains:wedge-S-formal}), within which the Gaussian factors in
$R(r,t)$ can be replaced by one.  Using the small-argument expansion of the
Bessel function, we then obtain
\begin{align}
\label{c-domains:wedge-S}
S(t) =\int_0^{\Theta}\!\!\! \sin(\nu\theta) \,d\theta\,
\int_0^\infty\!\!\! r\, R(r,t)\, dr \propto \int_0^{\sqrt{Dt}}\! \frac{1}{Dt}
\left(\frac{rr_0}{Dt}\right)^\nu \, r\, dr
  \sim \left(\frac{r_0}{\sqrt{Dt}}\right)^{\pi/\Theta}.
\end{align} 
The basic result is that the survival probability of a diffusing particle in
a wedge of opening angle $\Theta$ decays with time as
\begin{equation}
\label{c-domains:wedge-S-quote}
S(t)\sim t^{-\pi/2\Theta}\equiv t^{-\alpha}.
\end{equation}
The striking feature of this formula is that the exponent $\alpha$ depends on
the wedge opening angle in a non-trivial way, with $\alpha\to\infty$ as
$\Theta\to 0$ and $\alpha\to 1/4$ for $\Theta \to 2\pi$.

\subsection*{Conformal Transformations and Electrostatic Methods}
\label{sec:c-domains-conf}

Let's now solve the same wedge problem by exploiting conformal
transformations, together with the connection between first passage and
electrostatics.  To set the stage for the wedge geometry, consider the
first-passage probability for a diffusing particle in two dimensions to an
absorbing infinite line.  This problem may also be solved elegantly by the
electrostatic formulation.  In this approach, the time-integrated
concentration $\mathcal{C}(x,y)=\int_0^\infty c(x,y,t)\, dt$ obeys the Laplace
equation
\begin{align*}
  \nabla^2\mathcal{C}(z)= -\frac{1}{2\pi D}\; \delta(z-z_0)\,,
\end{align*}
where $z=(x,y)$ is the complex co-ordinate and the factor $1/(2\pi D)$
ensures the correct normalization.  Using the image method for
two-dimensional electrostatics, we find that the complex potential is
\begin{equation}
  \label{c-domains:imagesoln}
  \mathcal{C}(z)= \frac{1}{2\pi D}\ln\frac{z-z_0}{z- z_0^*}
=\frac{1}{2\pi D}\ln\frac{x-x_0+i(y-y_0)}{x-x_0+i(y+y_0)},
\end{equation}
where the asterisk denotes complex conjugation.  Finally, the time-integrated
flux that is absorbed at $x$ coincides with the electric field at this point.
This is
\begin{align}
  \label{c-domains:Eelec}
  {\mathcal{E}}(x|x_0,y_0) &= - D\,\frac{\partial \mathcal{C}(x,y)}{\partial y}\bigg|_{y=0}\nonumber \\
    &= \frac{1}{2\pi}\left(\frac{1}{y_0+i(x-x_0)} + \frac{1}{y_0-i(x-x_0)}\right) \nonumber\\
    &= \frac{1}{\pi} \frac{y_0}{(x-x_0)^2+y_0^2}\,.
\end{align}

We now use a conformal transformation to extend the result for the hitting
probability to the infinite line to the hitting probability in the wedge.
Consider the transformation $w=f(z)=z^{\pi/\Theta}$ that maps the interior of
the wedge of opening angle $\Theta$ to the upper half plane.  In complex
co-ordinates, the electrostatic potential in the wedge is
\begin{equation}
\label{c-domains:wedge-phi}
\mathcal{C}(z)= \frac{1}{2\pi D}\ln \frac{z^{\pi/\Theta}-z_0^{\pi/\Theta}}
{z^{\pi/\Theta}- (z_0^*)^{\pi/\Theta}}.
\end{equation}
From this expression and using the analogy between electrostatics and first
passage, we can extract time-integrated first-passage properties in the
wedge.  For example, the probability of being absorbed at a distance $x$ from
the wedge apex, when a particle begins at a unit distance from the apex along
the wedge bisector, is just the electric field at this point
\begin{eqnarray}
  \label{c-domains:Eelec-gen}
  {\mathcal{E}}(x|x_0,y_0)= \frac{1}{\Theta}
  \frac{x^{\pi/\Theta-1}}{1+x^{2\pi/\Theta}}\to x^{-(1+\pi/\Theta)} \qquad
  \text{for~} x\to\infty\,.
\end{eqnarray}

Although the electrostatic formulation ostensibly gives only time-integrated
first-passage properties, we can adapt it to also give time-dependent
features.  This adaptation is based on the following re-interpretation of the
equivalence between electrostatics and diffusion: an electrostatic system
with a point charge and specified boundary conditions is identical to a
diffusive system in the same geometry and boundary conditions, in which a
continuous source of particles is fed in at the location of the charge
starting at time $t=-\infty$.  Suppose now that the particle source is
``turned on'' at $t=0$.  Then, in a near zone that extends out to a distance
of the order of $\sqrt{Dt}$ from the source, the concentration has sufficient
time to reach its steady-state value.  Within this zone, the diffusive
solution converges to the Laplacian solution.  Outside this zone, however,
the concentration is close to zero.  This almost-Laplacian solution provides
the time integral of the survival probability up to time $t$.  We can then
deduce the survival probability by differentiating the concentration that is
due to this finite-duration source.

Thus suppose that a constant source of diffusing particles at $z_0$ inside
the absorbing wedge is turned on at $t=0$.  Within the region where the
concentration has had time to reach the steady state, $|z_0|<|z|<\sqrt{Dt}$,
the density profile is approximately equal to the Laplacian solution,
$\mathcal{C}(z)\sim |z|^{-\pi/\Theta}$.  We can neglect the angular dependence of
$\mathcal{C}(z)$ in this zone, as this dependence is immaterial for the survival
probability.  Conversely, for $|z|>\sqrt{Dt}$, the particle concentration is
vanishingly small because a particle is unlikely to diffuse such a large
distance.  From the analogy between electrostatics and first passage, the
near-zone density profile is just the same as the time integral of the
diffusive concentration.  Thus, by using the equivalence between the spatial
integral of this near-zone concentration in the wedge and the time integral
of the survival probability, we have
\begin{equation}
\label{c-domains:int-S}
\int_0^t S(t')\, dt' \approx \int_0^{\sqrt{Dt}} r^{-\pi/\Theta}\, r\,dr \sim
t^{1-\pi/2\Theta}.
\end{equation}
Since the total density injected into the system equals $t$, the survival
probability in the wedge is roughly
$\int_0^t S(t')\,dt'/t\sim t^{1-\pi/2\Theta}/t$, which gives
$S(t)\sim t^{-\pi/2\Theta}$.

\section{Stochastic Hunting in One Dimension}
\label{sec:hunting}

What is the time dependence of the survival probability of a diffusing lamb
that is hunted by $N$ diffusing lions?  We define this survival probability
as $S_N(t)$.  This toy problem is most interesting in a one-dimensional
geometry where all the lions are located to one side of the lamb.  It is
known~\cite{BG91} that this survival probability asymptotically decays
algebraically with time,
\begin{align}
  S_N(t)\sim t^{-\beta_N}\,,
\end{align}
and the goal is to compute the decay exponent $\beta_N$.  As we shall
discuss, the decay exponent is known for $N=1$ and $N=2$ only:
$\beta_1=\frac{1}{2}$, and $\beta_2=\frac{3}{4}$.  For $N\geq 3$, $\beta_{N}$
grows slowly with $N$, and numerical simulations give $\beta_3\approx 0.91$,
$\beta_4\approx 1.03$, and $\beta_{10}\approx 1.4$.  The focus of this
section is to derive $\beta_2=\frac{3}{4}$ by a simple geometric approach and
to develop some analytical understanding of the dependence of $\beta_N$ on
$N$ for $N\to\infty$.

Let us begin by treating a lamb that starts at $x_0>0$ and a single lion that
starts at $x=0$.  For simplicity, the diffusivities of the lamb and the lion
are assumed to both equal $D$.  The separation between the lamb and the lion
thus diffuses with diffusion coefficient $2D$.  When this separation reaches
zero, the lamb has been eaten.  This problem is just the classic
first-passage problem on the positive infinite line, except that the
diffusion coefficient is $2D$.  The probability that the lamb survives until
time $t$ is the same as the first-passage time being greater than $t$.  This
probability therefore is
\begin{align}
 \label{Svst}
S_1(t) &= \int_t^\infty \! \frac{x_0}{\sqrt{8\pi Dt'^3}} \;
e^{-x_0^2/8D t'}\, dt'= \text{erf}\biggl(\frac{x_0}{\sqrt{8D t}}\biggr)
 \sim \frac{x_0}{\sqrt{8\pi D t}}\,.
\end{align}
Thus the survival probability of the lamb asymptotically decays as
$t^{-1/2}$.  While the lamb is sure to die, its average lifetime is infinite.
Thus a single diffusing lion is not a particularly good hunter and it might
starve before eating the lamb.

\begin{figure}[ht]
  \begin{center}
    \subfigure[]{\includegraphics[width=0.475\textwidth]{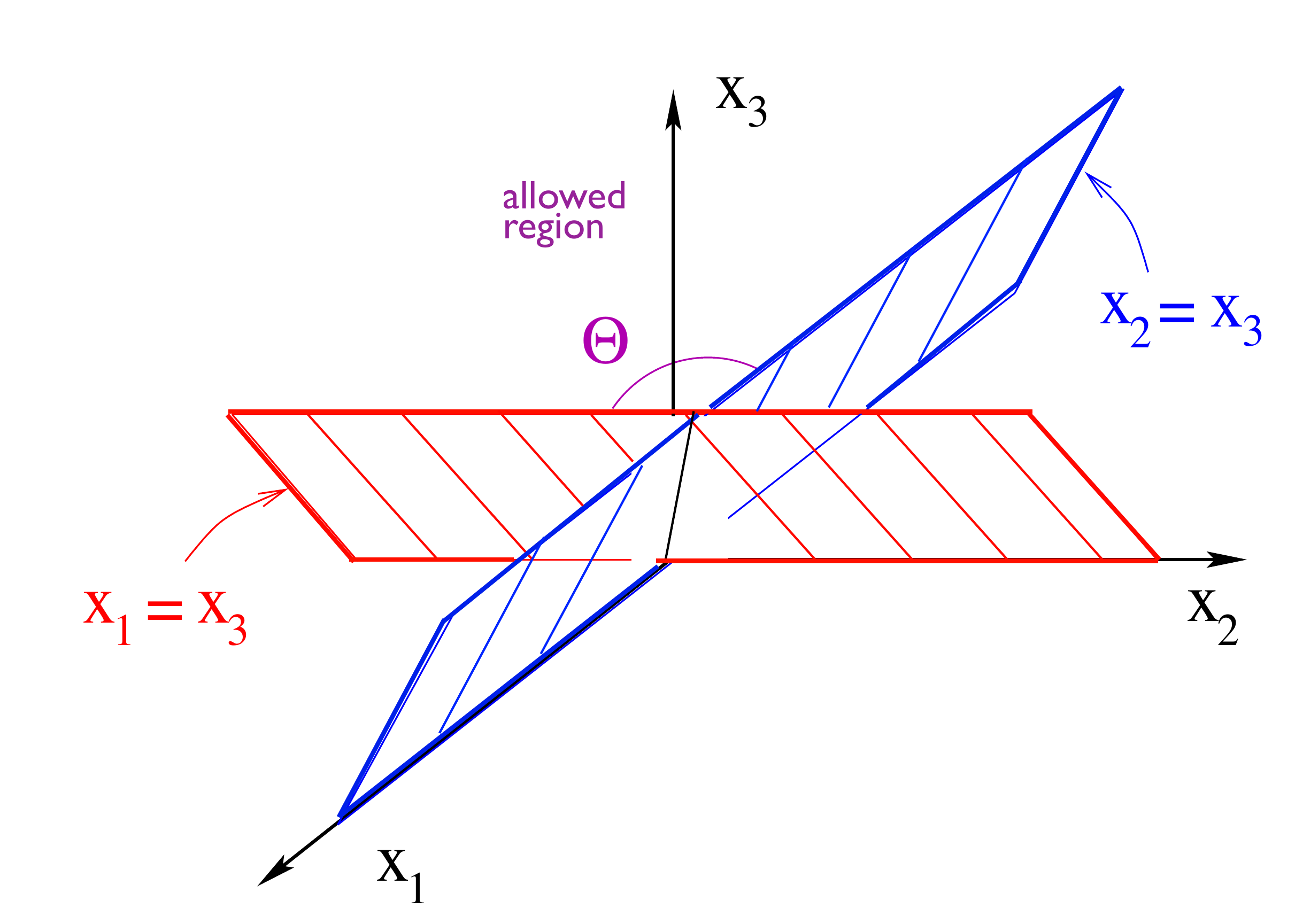}}\qquad\qquad
    \subfigure[]{\includegraphics[width=0.4\textwidth]{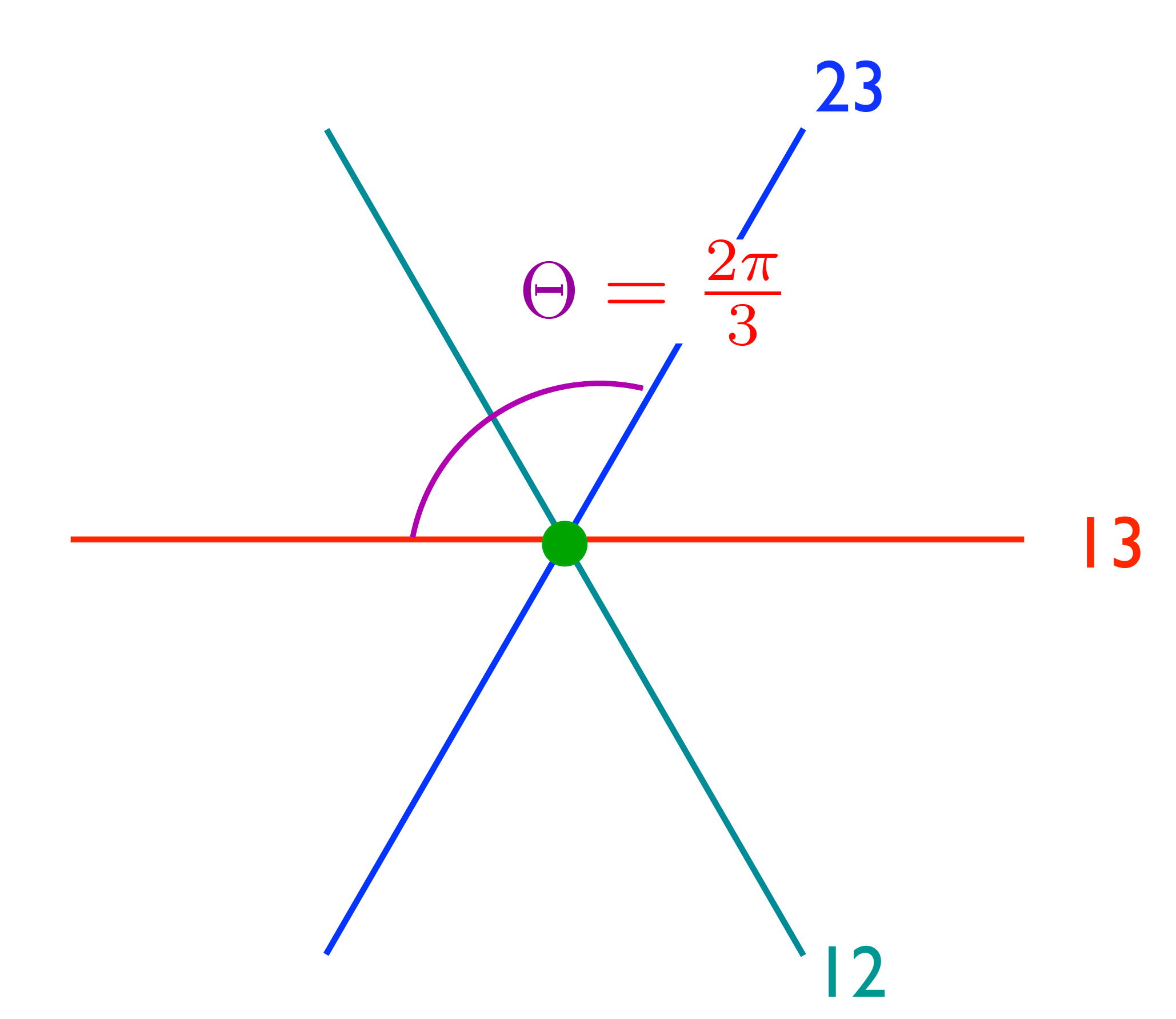}}
 \caption{(a) The allowed region for the effective particle that corresponds
   to the motion of the lamb and two lions.  (b) A view of the constraint
   planes perpendicular to the (1,1,1) axis.  The constraint plane $x_1=x_2$
   is also shown}~\label{fig:wedges}
  \end{center}
\end{figure}

What happens when there are $N=2$ lions?  We again assume that the lions
start from the origin while the lamb starts at $x_0>0$, and that the
diffusivities of all particles are the same.  Let us label the positions of
the lions as $x_1$ and $x_2$, and the position of the lamb as $x_3$ The lamb
survives up to time $t$ if the conditions $x_3>x_2$ and $x_3>x_1$ always
hold.  We can give an insightful geometric interpretation of this problem by
viewing the motion of the three particles on the line as equivalent to the
motion of a single effective particle in three dimensions with coordinates
$(x_1,x_2,x_3)$.  The constraints $x_3>x_2$ and $x_3>x_1$ mean that the
effective particle in three-space remains to the left of the plane $x_2=x_3$
and behind the plane $x_1=x_3$ (Fig.~\ref{fig:wedges}(a)).  This allowed
region is a wedge of opening angle $\Theta$ that is defined by the
intersection of these two planes. If the particle hits one of the planes,
then one of the lions has eaten the lamb.

This mapping therefore provides the lamb survival probability, since the
survival probability of a diffusing particle within this absorbing wedge
asymptotically decays as $S_{\rm wedge}(t)\sim t^{-\pi/2\Theta}$.  What is
the opening angle of this wedge?  We can determine this angle in a simple way
by also including the plane $x_1=x_2$ and then viewing the system along the
$(1,1,1)$ axis (Fig.~\ref{fig:wedges}(b)).  It is then clear that the wedge
angle is $2\pi/3$.  Substituting this result in
Eq.~\eqref{c-domains:wedge-S-quote}, we find that $\beta_2=\frac{3}{4}$.
Notice that $S_2(t)>S_1(t)^2$.  This inequality reflects the fact that the
incremental threat to the lamb from the second lion is less than the first.

In general, we can map the motion of the lamb and $N$ lions in one dimension
to a single effective particle in $N+1$ dimensions, with absorption when the
effective particle hits any of the $N$ constraint planes.  However, the
calculation of the survival probability of the effective particle within the
domain where the effective particle is confined---known as a Weyl
chamber---appears to be intractable.

\begin{figure}[ht]
  \begin{center}
\includegraphics[width=0.5\textwidth]{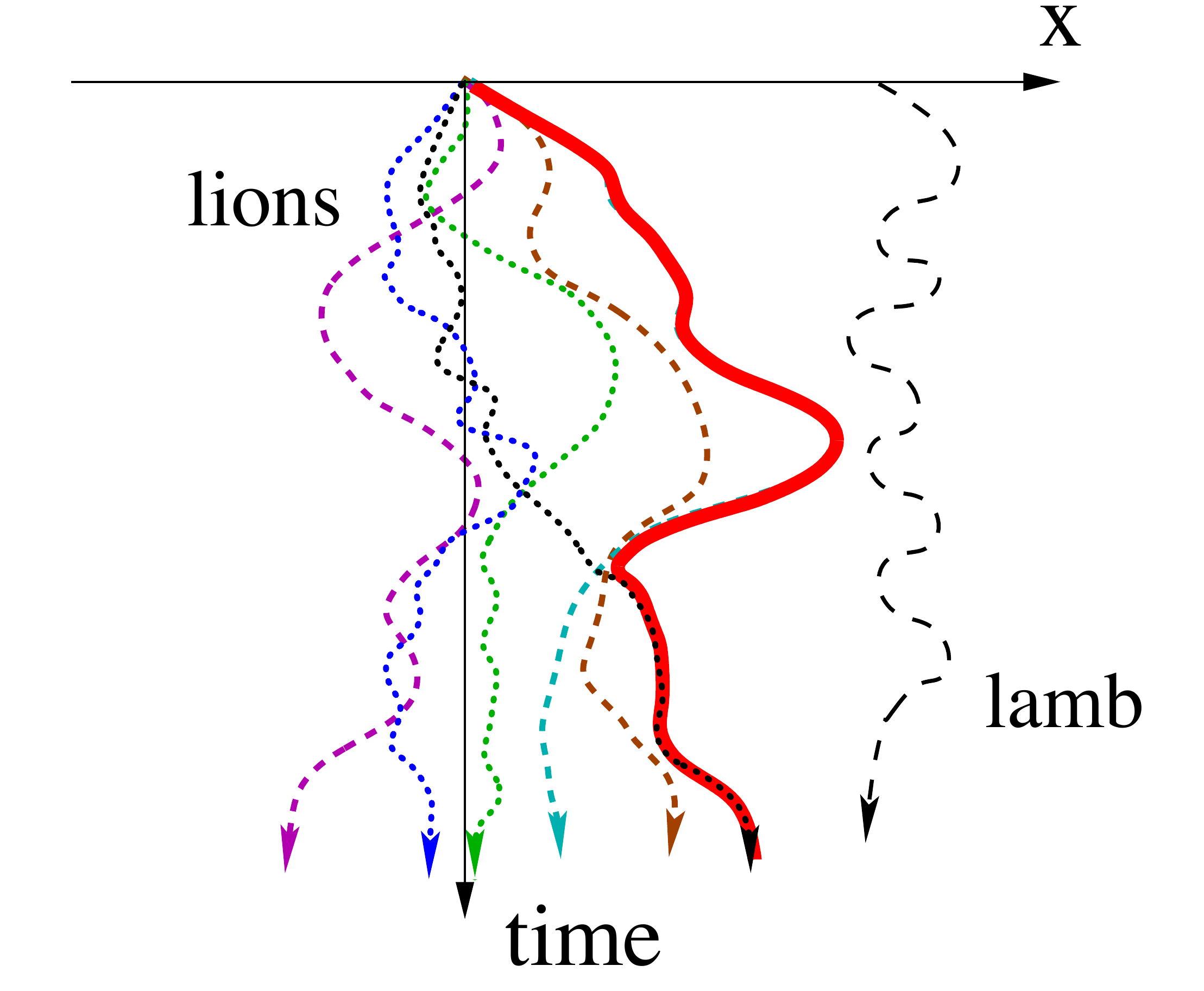}
\caption{Space-time trajectories of 6 lions and 1 lamb.  Although the motion
  of each lion is isotropic, the trajectory of the lion (whose identity may
  change) that happens to be closest to the lamb (solid red curve) tends to
  move towards the lamb.}~\label{fig:space-time}
  \end{center}
\end{figure}

\begin{figure}[ht]
  \begin{center}
\includegraphics[width=0.65\textwidth]{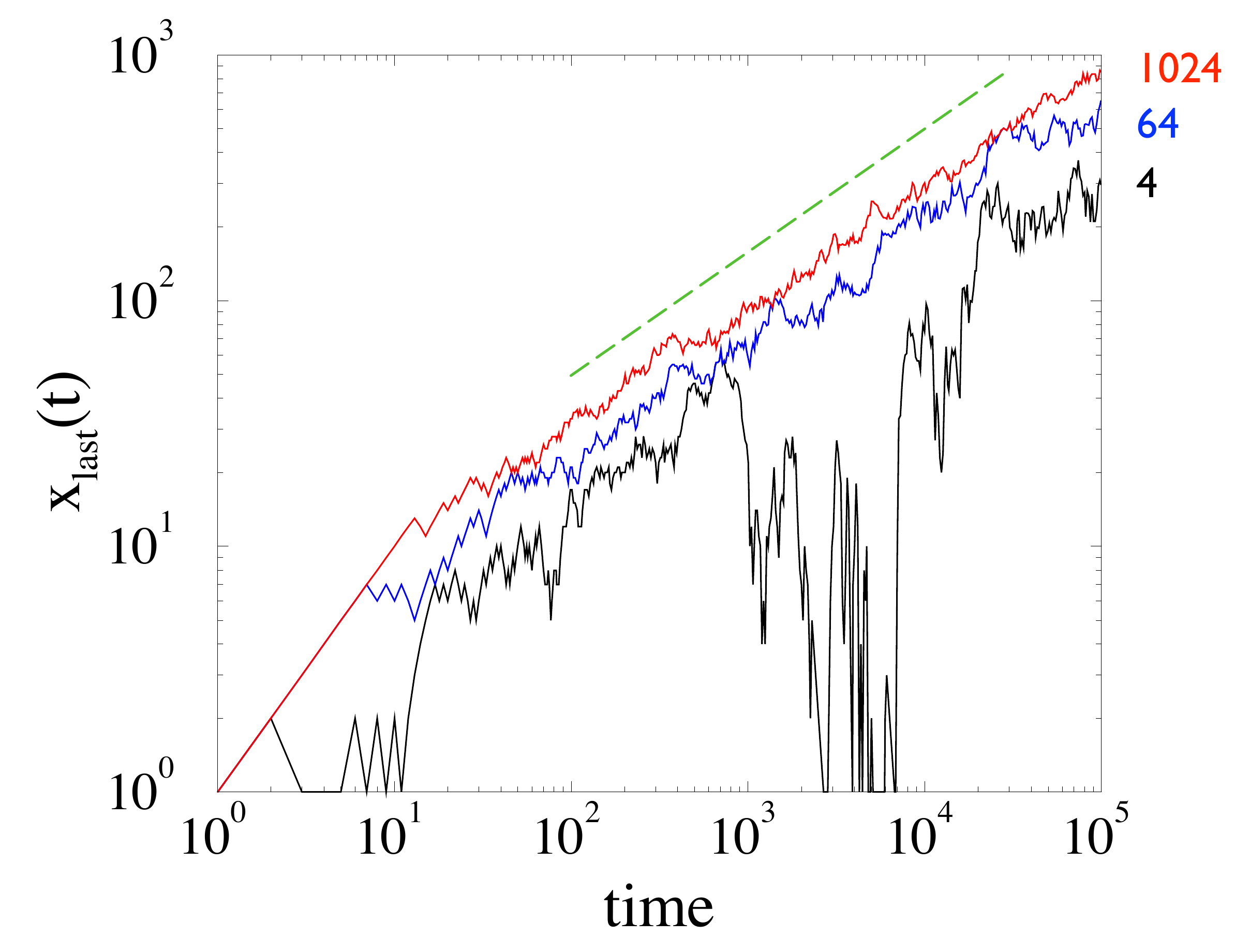}
\caption{Position of the last lion versus time for 4, 64, and 1024 lions.}~\label{fig:xlast}
  \end{center}
\end{figure}

While the problem for $N=3,4,\ldots$ lions is difficult, the problem becomes
much simpler for large $N$.  To determine the lamb survival probability for
large $N$, we only need to focus on the lion closest to the lamb, because
this last lion ultimately kills the lamb. As shown in
Fig.~\ref{fig:space-time}, the individual identity of this last lion can
change with time due to the crossing of different lion trajectories.  For
large $N$, there is a systematic bias of the motion of the last lion,
$x_{\rm last}(t)$.  This bias becomes stronger for increasing $N$, so that
$x_{\rm last}(t)$ becomes smoother as $N$ increases (Fig.~\ref{fig:xlast}).
This gradual approach of the last lion to the lamb is the mechanism which
leads to the survival probability of the lamb decaying as $t^{-\beta_N}$,
with $\beta_N$ a slowly increasing function of $N$.

To estimate the location of this last lion when $N\gg 1$ lions are initially
at the origin, we use the extreme statistics condition~\cite{G87}
\begin{equation}
\label{xlast-defn}
\int_{x_{\rm last}}^\infty \frac{N}{\sqrt{4\pi Dt}}\,e^{-x^2/4Dt}\,\,dx= 1.
\end{equation}
Equation~(\ref{xlast-defn}) states that one lion out of an initial
group of $N$ is in the range $[x_{\rm last},\infty]$.  Although the
integral in Eq.~(\ref{xlast-defn}) can be expressed in terms of the
complementary error function, it is instructive to evaluate it
approximately in a self consistent way by writing
$x=x_{\rm last}+\epsilon$ and re-expressing the integrand in terms of
$\epsilon$.  We thus find
\begin{equation*}
%\label{xlast-app}
\int_0^\infty \frac{N}{\sqrt{4\pi Dt}}\,e^{-x_{\rm last}^2/4Dt}\,
e^{-x_{\rm last}\epsilon/2Dt}\, e^{-\epsilon^2/4Dt}
\,\,d\epsilon= 1.
\end{equation*}
Now the second term in the integrand,
\begin{align*}
  e^{-x_{\rm last}\epsilon/2Dt}\,,
\end{align*}
is non-negligible for $\epsilon<2Dt/x_{\rm last}$.  Over this range of
$\epsilon$, the third exponential factor is of the order of
\begin{align*}
  e^{-Dt/x_{\rm last}^2}\,.
\end{align*}
If we use the result for $x_{\rm last}$ in Eq.~\eqref{xl-N}, the
above exponential factor becomes
\begin{align*}
  e^{-1/(4\ln N)}\,,
\end{align*}
which is very close to 1 for large N. If we thus ignore this term, the
integral above reduces to simple exponential decay, with the result
\begin{equation}
\label{xlast-result}
\frac{N}{\sqrt{4\pi Dt}}\,e^{-x_{\rm last}^2/4Dt}\,\frac{2Dt}{x_{\rm last}}= 1.
\end{equation}

We now define $y=x_{\rm last}/\sqrt{4Dt}$ and $M=N/\sqrt{4\pi}$, so that the
above condition can be simplified to $y\exp(e^{y^2})=M$, whose asymptotic
solution is
\begin{equation*}
%\label{xlast-final}
y=\sqrt{\ln M\,}\Bigl(1-\frac{1}{4}\frac{\ln (\ln M)}{\ln
M}+\ldots\Bigr).
\end{equation*}
To lowest order, this gives
\begin{equation}
\label{xl-N}
x_{\rm last}(t)\sim\sqrt{4D\ln N\,t}\equiv\sqrt{A_N t},
\end{equation}
for $N$ finite. For $N=\infty$, $x_{\rm last}(t)$ would always equal $t$ if
an infinite number of discrete random walking lions were initially at the
origin. A more suitable initial condition therefore is a concentration $c_0$
of lions that are uniformly distributed from $-\infty$ to 0. In this case,
only $N\propto\sqrt{c^2_0 D t}$ of the lions are ``dangerous,'' that is,
within a diffusion distance from the edge of the pack and thus potential
candidates for eating the lamb.  Consequently, for $N\to\infty$, the leading
behavior of $x_{\rm last}(t)$ becomes
\begin{equation}
\label{xl-inf}
x_{\rm last}(t) \sim \sqrt{2D\ln(c^2_0 D t)\,t}\, .
\end{equation}

An important feature of the time dependence of $x_{\rm last}$ is that
fluctuations in this quantity decrease for large $N$ (Fig.~\ref{fig:xlast}).
Therefore the lamb and $N$ diffusing lions can be recast as a two-body system
of a lamb and an absorbing boundary which {\em deterministically} advances
toward the lamb according to $x_{\rm last}(t)=\sqrt{A_Nt}$.  This determinism
is what makes the problem for large $N$ tractable.

To solve this effective two-body problem, it is convenient to change
coordinates from $(x,t)$ to $('=x-x_{\rm last}(t),t)$ to fix the absorbing
boundary at the origin. By this construction, the diffusion equation for the
lamb probability distribution is transformed to the convection-diffusion
equation
\begin{equation}
\label{CDE}
\frac{\partial p(x',t)}{\partial t}- \frac{x_{\rm last}}{2t}
\frac{\partial p(x',t)}{\partial x'}= D\frac{\partial^2 p(x',t)}{\partial x'^2}\,, 
\quad (0\le x' <\infty)
\end{equation}
with the absorbing boundary condition $p(x'=0,t)=0$.  In this reference frame
that is fixed on the average position of the last lion, the second term in
Eq.~(\ref{CDE}) accounts for the bias of the lamb towards the absorber with
an effective speed $-x_{\rm last}/2t$. Because $x_{\rm last}\sim \sqrt{A_Nt}$
and $x'\sim\sqrt{D t}$ have the same time dependence, the lamb survival
probability acquires a nontrivial dependence on the dimensionless parameter
$A_N/D$.  This behavior arises whenever there is a coincidence of fundamental
length scales in the system (see, e.g., \cite{B96} for other such examples).

Equation~(\ref{CDE}) can be transformed into the parabolic cylinder equation
by first introducing the dimensionless length $\xi=x'/x_{\rm last}$ and
making the following scaling ansatz for the lamb probability density,
\begin{equation}
\label{cxt}
p(x',t) \sim t^{-\beta_N-1/2}\mathcal{C}(\xi).
\end{equation}
The power law prefactor in Eq.~(\ref{cxt}) ensures that the integral of
$p(x',t)$ over all space, namely the survival probability, decays as
$t^{-\beta_N}$, and $\mathcal{C}(\xi)$ expresses the spatial dependence of the lamb
probability distribution in scaled length units. This ansatz codifies the
fact that the probability density is not a function of $x'$ and $t$
separately, but is a function only of the dimensionless ratio $x'/x_{\rm
 last}(t)$. 

Substituting Eq.~(\ref{cxt}) into Eq.~(\ref{CDE}), we obtain
\begin{equation}
\label{de-scaled}
\frac{D}{A_N}\frac{d^2{\mathcal{C}}}{d \xi^2}+\frac{1}{2}(\xi+1)\frac{d\mathcal{C}}{d\xi}+
\bigl(\beta_N+\frac{1}{2}\bigr)\mathcal{C}=0.
\end{equation}
By introducing $\eta=(\xi+1)\sqrt{A_N/2D}$ and
$\mathcal{C}(\xi)=e^{-\eta^2/4}\,{\mathcal{D}}(\eta)$ in Eq.~(\ref{de-scaled}), we are led to
the parabolic cylinder equation of order $2\beta_N$~\cite{BO78}
\begin{equation}
\label{pce}
\frac{d^2{\mathcal{D}}_{2\beta_N}}{d\eta^2}+
\biggl[2\beta_N+\frac{1}{2}-\frac{\eta^2}{4}\biggr]{\mathcal{D}}_{2\beta_N}=0,
\end{equation}
subject to the boundary condition, ${\mathcal{D}}_{2\beta_N}(\eta)=0$ for both
$\eta=\sqrt{A_N/2D}$ and $\eta=\infty$. Equation~(\ref{pce}) has the
form of a Schr\"odinger equation for a quantum particle of energy
$2\beta_N+\frac{1}{2}$ in a harmonic oscillator potential $\eta^2/4$ for
$\eta>\sqrt{A_N/2D}$, but with an infinite barrier at
$\eta=\sqrt{A_N/D}$~\cite{S68}. For the long-time behavior, we need to
find the ground state energy in this potential. For $N\gg 1$, we may
approximate this energy as the potential at the classical turning point, that
is, $2\beta_N+\frac{1}{2}\simeq \eta^2/4$. We therefore obtain
$\beta_N\sim A_N/16D$. Using the value of $A_N$ given in Eqs.~(\ref{xl-N})
and (\ref{xl-inf}) the decay exponent is
\begin{equation}
\label{beta-N}
\beta_N\sim
\begin{cases} \tfrac{1}{4} \ln N &  N~ \text{finite}\\[2mm]
  \tfrac{1}{8} \ln t &  N=\infty\,.
\end{cases}
\end{equation}
The latter dependence of $\beta_N$ implies that for $N\to\infty$,
the survival probability has the log-normal form
\begin{equation}
\label{S-inf}
S_{\infty}(t)\sim \exp\left(-\tfrac{1}{8}\,\ln^2t\right).
\end{equation}
The important feature of the exponent $\beta_N$ is its very slow
increase with $N$.  That is, each successive lion that is added to the
hunt has a decreasing influence on the survival of the lamb.  Indeed,
only a small subset of the lions for large $N$ actually have a chance
to catch the lamb.

\section{The Expanding Interval}
\label{sec:growing-interval}

We have seen that the survival probability $S(t)$ of a diffusing particle in
a fixed-length absorbing interval of length $L$ asymptotically decays as
$e^{-\pi^2Dt/L^2}$.  What happens if the interval length grows with time,
$L(t)\sim t^\alpha$?  This simple question illustrates the relative effects
of diffusion and the motion of the boundary on first-passage properties.
This interplay is a classic problem in the first-passage literature,
especially when the boundary motion matches that of diffusion.  Solutions to
this problem have been obtained by a variety of methods (see, e.g.,
\cite{B66,D69,U80,S88}).  Here we give a physics-based approach that is based
on \cite{KR96}.

It is easy to infer the survival probability for a slowly expanding interval.
Here, slowly expanding means that the interval length grows slower than
diffusion.  Consequently, the probability distribution of the particle
spreads faster than the interval grows and so that the survival probability
should decay rapidly with time.  Using an adiabatic approximation that one
typically encounters in basic quantum mechanics, we will show that
$S(t)\sim\exp\big[-At^{(1-2\alpha)}\big]$, where $A$ is a constant.  Conversely,
for the rapidly expanding interval, $\alpha>1/2$, the particle is unlikely
reach the either end of the interval and the probability distribution is
close to that for free diffusion.  This is the basis of the \emph{free
  approximation} that leads to a non-zero limiting value for $S(t)$ as
$t\to\infty$.

In the marginal case where the interval expands at the same rate as
diffusion, $L(t) = \sqrt{At}$, a new dimensionless parameter arises---the
ratio of the diffusion length to the interval length.  As we shall show, this
leads to $S(t)$ decaying as a non-universal power-law in time,
$S(t)\sim t^{-\beta}$, with $\beta=\b(A,D)$ diverging for $A/D\ll 1$ and
approaching zero for $A/D\gg 1$.

\subsection*{Slowly Expanding Interval}

For $L(t)\ll\sqrt{Dt}$, we invoke the adiabatic approximation~\cite{S68}, in
which the spatial dependence of the concentration for an interval of length
$L(t)$ is assumed to be identical to that of the static diffusion equation at
the instantaneous value of $L$.  This assumption is based on the expectation
that the concentration in a slowly expanding interval is always close to that
of a fixed-size interval.  Thus we write
\begin{equation}
\label{c-examples:cage1}
c(x,t)\simeq f(t)\,\cos\left(\frac{\pi x}{2L(t)}\right)\equiv c_{\rm ad}(x,t),
\end{equation}
with $f(t)$ to be determined.  The corresponding survival probability is
\begin{equation}
\label{c-examples:cage4}
S(t)\approx\int_{-L(t)}^{L(t)} ~c_{\rm ad}(x,t)\,dx = \frac{4}{\pi}~f(t)L(t).
\end{equation}
For convenience, we now define the interval boundaries as $[-L(t),L(t)]$.  To
obtain $f(t)$, we substitute approximation~(\ref{c-examples:cage1}) into the
diffusion equation, as in separation of variables, to give
\begin{equation}
\label{c-examples:cage2}
\frac{df}{dt} = -\left(\frac{D\pi^2}{4L^2}\right) f - \left(\frac{\pi x}{2L^2}\right)\,
\frac{dL}{dt}\, \tan\left(\frac{\pi x}{2L}\right)\,  f.
\end{equation}
Notice that variable separation does not strictly hold, since the equation
for $f(t)$ also involves $x$.  However, when $L(t)$ increases as $(At)^\alpha$
with $\alpha<1/2$, the second term on the right-hand side is negligible.  Thus we
drop this second term and solve the simplified form of
\eqref{c-examples:cage2}.  We thereby find that the controlling factor of
$f(t)$ is given by
\begin{eqnarray}
\label{c-examples:cage3}
f(t) \sim \exp\left[-\frac{D\pi^2}{4} \int_0^t\,\frac{dt'}{L^2(t')}\right]
= \exp\left[-\frac{D\pi^2}{4(1-2\alpha)A^{2\alpha}}\,t^{1-2\alpha}\right].
\end{eqnarray}
Notice that $f(t)$ reduces to a pure exponential decay for a fixed-length
interval, while for $\alpha\to 1/2$, Eq.~\eqref{c-examples:cage3} suggests a
more slowly decaying functional form for $S(t)$.

\subsection*{Rapidly Expanding Interval}

For a rapidly expanding interval, the escape rate from the system is small
and the absorbing boundaries should eventually become irrelevant.  We
therefore expect that the concentration profile should approach the Gaussian
distribution of free diffusion at long times~\cite{KR96}.  We may then
account for the slow decay of the survival probability by augmenting the
Gaussian with an overall decaying amplitude.  This free approximation is a
nice example in which the existence of widely separated time scales,
$\sqrt{Dt}$ and $L(t)$, suggests the nature of the approximation itself.

According to the free approximation, we write
\begin{align*}
c(x,t)  \approx \frac{S(t)}{\sqrt{4\pi Dt}}\,e^{-x^2/4Dt}\equiv c_{\rm free}(x,t)\,.
\end{align*}
Although this concentration does not satisfy the absorbing boundary
condition, the inconsistency is negligible at large times, since the density
is exponentially small at the interval boundaries.  We may now find the time
dependence of the survival probability by equating the probability flux to
the interval boundaries, $2D|\frac{\partial c}{\partial x}|$, to the loss of
probability within the interval.  For $L(t)=(At)^\alpha$, this flux is
\begin{equation}
\label{c-examples:cage-flux}
\frac{S(t) A^\alpha}{\sqrt{4\pi D}}\;t^{\alpha-3/2}\,
\exp\left(-\frac{A^{2\alpha}}{4D}t^{2\alpha-1}\right),
\end{equation}
which rapidly goes to zero for $\alpha>1/2$.  Since this flux equals
$-\frac{dS}{dt}$, it follows that the survival probability approaches
a non-zero limiting value for $\alpha>1/2$, and that this limiting
value goes to zero as $\alpha\to 1/2$.  Explicitly,
\begin{align}
  \frac{dS(t)}{dt} = - \frac{1}{S(t)}\;B\, t^{\alpha-3/2}\; \exp\left(- C\, t^{2\alpha-1}\right)\,,
\end{align}
where $B\equiv A^\alpha/\sqrt{4\pi D}$ and $C\equiv A^{2\alpha}/(4D)$.
We now introduce $y= C t^{2\alpha-1}$ and change the integration
variable from $t$ to $y$.  After some straightforward steps we have
\begin{align}
  \ln S(t\!=\!\infty) &=-\frac{B}{2\alpha-1} \int_0^\infty \left(\frac{y}{C}\right)^{1/2} e^{-y}\; \frac{dy}{y}\nonumber\\
            &=- \frac{B}{C^{1/2}}\frac{1}{2\alpha-1} \,\Gamma\left(\tfrac{1}{2}\right)\,,
\end{align}
where $\Gamma(\cdot)$ is the Euler gamma function.  Thus a diffusing
particle has a non-zero probability to survive forever when the
interval grows fast enough.  This ultimate survival probability
rapidly goes to zero as $\alpha\to 1/2$ from above.

\subsection*{Marginally Expanding Interval}

For the marginal case of $\alpha=1/2$, the adiabatic and the free approximations
are ostensibly no longer appropriate, since $L(t)=\sqrt{At}$ and $\sqrt{Dt}$
have a fixed ratio.  However, for $A/D\ll 1$ and $A/D\gg 1$, we might hope
that these methods could still be useful.  Thus we continue to apply these
heuristic approximations in their respective domains of validity, $A/D\ll 1$
and $A/D\gg 1$, and check their accuracies {\it a posteriori}.  We will see
that the survival probability exponents predicted by these two approximations
are each quite close to the exact result except for $A/D\approx 1$.

When the adiabatic approximation is applied, the second term in
Eq.~(\ref{c-examples:cage2}) is, in principle, non-negligible for $\alpha=1/2$.
However, for $A\ll D$, the interval still expands more slowly (in amplitude) than
free diffusion and the error made by neglecting the second term in
Eq.~(\ref{c-examples:cage2}) may still be small.  The solution to this
crudely truncated equation immediately gives $f(t)$, which, when substituted
into approximately~(\ref{c-examples:cage4}) leads to
$S_{\rm ad}(t) \approx t^{-\beta_{\rm ad}}$, with
\begin{equation}
\label{c-examples:cage6}
\beta_{\rm ad}\approx \frac{D \pi^2}{4A}-\frac{1}{2}.
\end{equation}
The trailing factor of $-1/2$ should not be taken very seriously, because the
neglected term in Eq.~(\ref{c-examples:cage2}) leads to additional
corrections to $\beta_{\rm ad}$ that are also of the order of 1.

\begin{figure}[ht]
  \begin{center}
    \includegraphics[width=0.65\textwidth]{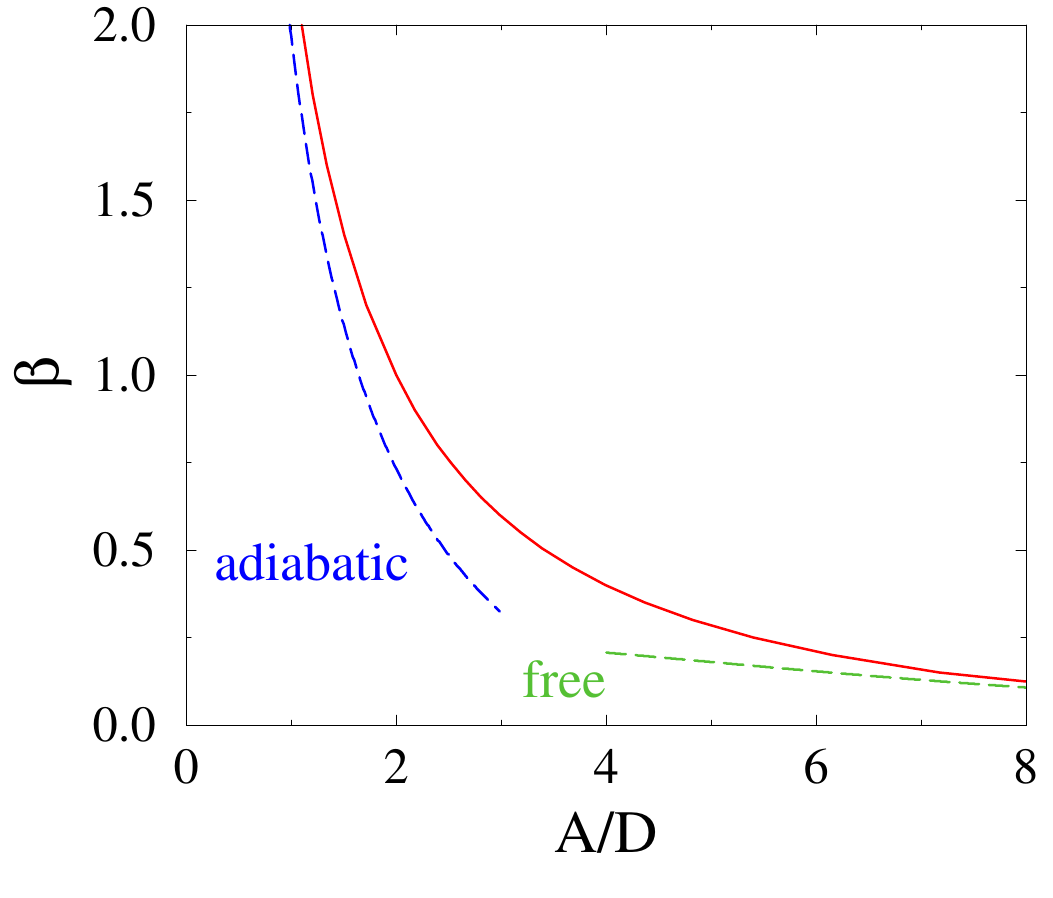}
    \caption{The survival probability exponent $\beta$ for the marginal case
      $L(t)=(At)^{1/2}$.  Shown is the numerically exact value of $\beta$
      (red, solid) from solution of Eq.~(\ref{c-examples:cage16}), together
      with the predictions from the adiabatic and the free approximations,
      Eqs.(\ref{c-examples:cage6}) and (\ref{c-examples:cage10}) respectively
      (dashed curves).}~\label{fig:beta}
  \end{center}
\end{figure}

Similarly for $A\gg D$, the free approximation gives
\begin{eqnarray}
\label{c-examples:cage9}
\frac{dS}{dt}  \approx -2D\bigg|\frac{\partial  c(x,t)}{\partial x}\bigg|_{x=\sqrt{At}}
        \;=\, -\frac{S(t)}{t}\sqrt{ \frac{A}{4\pi D}}\, e^{-A/4D}.
\end{eqnarray}
This again leads to the non-universal power law for the survival probability,
$S_{\rm free}\sim t^{-\beta_{\rm free}}$, with
\begin{equation}
\label{c-examples:cage10}
\beta_{\rm free}=\sqrt{ \frac{A}{4\pi D}}\, e^{-A/4D}.
\end{equation}
As shown in Fig.~\ref{fig:beta}, these approximations are surprisingly
accurate over much of the range of $A/D$.

To complete our discussion, we outline a first-principles analysis for the
survival probability of a diffusing particle in a marginally expanding
interval~\cite{KR96}.  When $L(t)=\sqrt{At}$, a natural scaling hypothesis is to
write the density in terms of the two dimensionless variables
\begin{align*}
\xi\equiv \frac{x}{L(t)}\qquad{\rm and}\qquad \rho=\sqrt{\frac{A}{D}}\,.
\end{align*}
We now seek solutions for the concentration in the form,
\begin{equation}
\label{c-examples:cage12}
c(x,t) = t^{-\beta-1/2}\;\mathcal{C}_\rho(\xi),
\end{equation}
where $\mathcal{C}_\rho(\xi)$ is a two-variable scaling function that encodes the
spatial dependence.  The power law prefactor ensures that the survival
probability, namely, the spatial integral of $c(x,t)$, decays as $t^{-\beta}$,
as defined at the outset of this section.

After substituting Eq.~(\ref{c-examples:cage12}) into the diffusion equation,
the scaling function satisfies the ordinary differential equation
\begin{align*}
  \frac{1}{\rho^2}\frac{d^2\mathcal{C}}{d\xi^2} +\frac{\xi}{2}\frac{d\mathcal{C}}{d\xi}
  +\left(\beta+\frac{1}{2}\right)\mathcal{C}=0\,.
\end{align*}
Then by introducing $\eta=\xi\,\sqrt{\rho/2}$ and $\mathcal{C}(\xi)=
e^{-\eta^2/4}\,{\mathcal{D}}(\eta)$, we transform this into the parabolic cylinder
equation~\cite{AS72}
\begin{equation}
\label{c-examples:cage14}
\frac{d{\mathcal{D}}}{d\eta^2} + \left[2\beta+\frac{1}{2}-\frac{\eta^2}{4}\right]{\mathcal{D}}=0.  
\end{equation}
When the range of $\eta$ is unbounded, this equation has solutions for
quantized values of the energy eigenvalue $E=2\beta+\frac{1}{2}=\frac{1}{2}$,
$\frac{3}{2}$, $\frac{5}{2},\ldots$~\cite{S68}.

For our interval problem, the range of $\eta$ is restricted to
$|\eta|\leq \sqrt{A/2D}$.  In the equivalent quantum mechanical system, this
corresponds to a particle in a harmonic-oscillator potential for
$|\eta|<\sqrt{A/2D}$ and an infinite potential for $|\eta|>\sqrt{A/2D}$.  For
this geometry, a spatially symmetric solution to
Eq.~(\ref{c-examples:cage14}), appropriate for the long-time limit for an
arbitrary starting point, is
\begin{align*}
  {\mathcal{D}}(\eta)\equiv \tfrac{1}{2}\big[{\mathcal{D}}_{2\beta}(\eta)
  +{\mathcal{D}}_{2\beta}(-\eta)\big]\,, 
\end{align*}
where ${\mathcal{D}}_\nu(\eta)$ is the parabolic cylinder function of order $\nu$.
Finally, the relation between the decay exponent $\beta$ and $\sqrt{A/D}$ is
determined implicitly by the absorbing boundary condition, namely,
\begin{equation}
\label{c-examples:cage16}
{\mathcal{D}}_{2\beta}(\sqrt{A/2D})+{\mathcal{D}}_{2\beta}(-\sqrt{A/2D})=0.
\end{equation}

This condition for $\beta=\beta(A/D)$ simplifies in the limiting cases $A/D\ll 1$
and $A/D\gg 1$.  In the former, the exponent $\beta$ is large and the second two
terms in the brackets in Eq.~(\ref{c-examples:cage14}) can be neglected.
Equivalently, the physical range of $\eta$ is small, so that the potential
plays a negligible role.  The solution to this limiting free-particle
equation is just the cosine function, and the boundary condition immediately
gives the limiting expression of Eq.~(\ref{c-examples:cage6}), but without
the subdominant term of $-1/2$.  In the latter case of $A\gg D$, $\beta\to 0$
and Eq.~(\ref{c-examples:cage14}) approaches the Schr\"odinger equation for
the ground state of the harmonic oscillator.  In this case, a detailed
analysis of the differential equation reproduces the limiting exponent of
Eq.~(\ref{c-examples:cage10}) (see \cite{KR96} for details).  These provide
rigorous justification for the limiting values of the decay exponent $\beta$
which we obtained by heuristic means.

\subsection*{The Khintchine Iterated Logarithm Law}
In the marginal situation of $L(t)=(At)^{1/2}$, we have seen that the
survival probability $S(t)$ decays as a power law $t^{-\beta(A,D)}$, with
$\beta\to 0$ as $A/D\to\infty$.  This decay becomes progressively slower as
$A$ increases.  On the other hand, when $L(t)\propto t^{\alpha}$, with
$\alpha$ strictly greater than 1/2, the survival probability at infinite time
is greater than zero.  This leads to the following natural question: what is
the nature of the transition between certain death, defined as
$S(t\to\infty)=0$, and a non-zero survival survival probability,
$S(t\to\infty)>0$?

The answer to this question is surprisingly rich.  There is an infinite
sequence of transitions, where $L(t)$ acquires additional iterated
logarithmic time dependences, which define regimes where $S(t)$ assumes
progressively slower functional forms.  The first term in this series is
known as the Khintchine \emph{iterated logarithm law} (\cite{K24,F68}).
While the Khintchine law has been obtained by rigorous methods, we can also
obtain this intriguing result, as well as the infinite sequence of
transitions, with relatively little computation by the free approximation.

Because we anticipate that the transition between life and death occurs when
$L(t)$ grows slightly faster than $(At)^{1/2}$, we make the hypothesis that
$A\propto u(t)$, with $u(t)$ growing slower than a power law in $t$.  Now
that $L(t)$ increases more rapidly than the diffusion length $(Dt)^{1/2}$,
the free approximation should be asymptotically exact, since it already works
extremely well when $L(t)=(At)^{1/2}$ with $A$ large.  Within this
approximation, we rewrite Eq.~(\ref{c-examples:cage9}) as
\begin{equation}
\label{c-examples:free-approx}
\ln S(t) \sim  -\int^t \frac{dt'}{t'}\sqrt{\frac{A}{4\pi D}}\,e^{-A/4D}.
\end{equation}
Here we neglect the lower limit, since the free approximation is valid only
as $t\to\infty$, where the short-time behavior is irrelevant.  In this form,
it is clear that for $L=(At)^{1/2}$, $\ln S$ decreases by an infinite amount
for $t\to\infty$ because of the divergence of the integral.  Thus
$S(t\to\infty)\to 0$.  To make the integral converge, the other factors in
the integral must somehow cancel the logarithmic divergence that arises from
the factor $dt/t$.  Accordingly, let us substitute $L(t)=\sqrt{4Dt\,u(t)}$
into the approximation \eqref{c-examples:free-approx}.  This gives
\begin{equation*}
%\label{c-examples:free-approx-new}
\ln S(t) \sim  -\frac{1}{\sqrt{\pi}}\int^t \frac{dt'}{t'}\sqrt{u(t')}\,e^{-u(t')}.
\end{equation*}

To simplify this integral, it is helpful to define $x=\ln t$ so that
\begin{equation}
\label{c-examples:free-approx-u}
S(x) \sim  -\frac{1}{\sqrt{\pi}}\int^{\ln t} dx\,\sqrt{u(x)}\,e^{-u(x)}.
\end{equation}
To lowest order, it is clear that if we choose $u(x)=\lambda \ln x$ with
$\lambda>1$, the integral converges as $t\to\infty$.  Thus the asymptotic
survival probability is positive.  Conversely, for $\lambda\leq 1$, the
integral diverges and the particle surely dies.  In this latter case,
evaluation of the integral to lowest order gives
\begin{equation}
\label{c-examples:free-S-t}
S(t)\sim \exp\left[-\frac{(\ln t)^{1-\lambda}\sqrt{\lambda\ln\ln t}}
{\sqrt{\pi}(1-\lambda)}\right]\qquad \lambda <1.
\end{equation}
This decay is slower than any power law, but faster than any power of
logarithm, that is, $t^{-\beta}<S(t)<(\ln t)^{-\gamma}$ for $\beta\to 0$ and
$\gamma\to\infty$.  

What happens in the marginal case of $\lambda=1$?  Here we can refine the
criterion between life and death still further by incorporating into $u(x)$ a
correction that effectively cancels the subdominant factor $\sqrt{u(x)}$ in
Eq.~(\ref{c-examples:free-approx-u}).  We therefore define $u(x)$ such that
$e^{-u(x)}=1/x(\ln x)^\mu$.  Then in terms of $y=\ln x$,
Eq.~(\ref{c-examples:free-approx-u}) becomes
\begin{equation}
\label{c-examples:free-approx-y}
S(y) \sim  -\frac{1}{\sqrt{\pi}}\int^{\ln\ln t}
\frac{dy}{y^{\mu-1/2}}\left(1+\frac{\mu\ln y}{y}\right).
\end{equation}
This integral now converges for $\mu> 3/2$ and diverges for $\mu\leq 3/2$.
In the latter case, the survival probability now lies between the bounds
$(\ln t)^{-\gamma}<S(t)<(\ln \ln t)^{-\delta}$ for $\gamma\to 0$ and
$\delta\to\infty$.  At this level of approximation, we conclude that
when the cage length grows faster than 
\begin{subequations}
\begin{equation}a
\label{L-crit}
L^*(t)=\sqrt{4Dt\left(\ln\ln t+\tfrac{3}{2}\ln\ln\ln t+\ldots\right)}
\end{equation}
a diffusing particle has a non-zero asymptotic survival probability, while
for a interval that expands as $L^*$, there is an extremely slow decay of the
survival probability.

By incorporating successively finer corrections into $u(x)$ and following the
same logic that led to Eq.~(\ref{c-examples:free-approx-y}), an infinite
series of correction terms can be generated in the expression for $L^*(t)$.
By this approach, the ultimate life-death transition corresponds to an
ultra-slow decay in which $S(t)$ has the form
$S(t)\sim\lim_{n\to\infty}1/\ln_n t$, where $\ln_2 t\equiv \ln\ln t$ and
$\ln_n t\equiv \ln\ln_{n-1} t$.  It is remarkable that the physically motivated and
relatively naive free approximation can generate such an intricate solution.
As a final note, P. Erd\"os sharpened the result of Eq.~\eqref{L-crit}
considerably and found that $L^*(t)$ has the infinite series representation
\begin{equation}
L^*(t)=\sqrt{4Dt\left(\ln_2 t+\tfrac{3}{2}\ln_3 t+\ln_4 t+\ln_5 t+\ldots\right)}\,,
\end{equation}
\end{subequations}
in which only the coefficient of the term multiplying $\ln_3 t$ is different
than 1. 

\section{Birth-Death Dynamics}

As our last topic, we determine the kinetics of the \emph{birth-death}
process.  We imagine a collection of identical particles, each of which gives
birth to an identical offspring with rate $\lambda$, and each particle can
independently die with rate $\mu$.  The goal is to determine the time
dependence of the population size.  As one can easily imagine, this is a
classic model for a variety of biological processes and there is vast
literature on this general topic (see, e.g.,~\cite{K49,KRB10}).

The most interesting case physically is the symmetric situation of equal
birth and death rates for each particle, $\lambda=\mu$, so that the average
population is static.  For $\mu>\lambda$, the population size decreases as
$e^{-(\mu-\lambda)t}$, which quickly goes to zero.  In the opposite case, the
population grows exponentially with time and an additional mechanism is needed
to cut off this growth.  For $\lambda=\mu$, the average population is fixed,
but the time dependence of the distribution of the number of particle
exhibits non-trivial kinetics on the positive infinite line.  We can
alternatively view the birth-death process as a continuous-time random walk
on the line, but with birth and death rates for the entire population that
are linear functions of $n$.  That is, the overall process is symmetric but
moves faster for a larger population.

Let $n$ denote the number of particles in the population.  The time
dependence of the average number of particles obeys the rate equation
$\langle\dot n\rangle = (\lambda-\mu)\langle n\rangle = 0$, where the
overdot denotes the time derivative.  Thus the average number of
particles is conserved, as is clear from the condition $\lambda=\mu$.
That is, the birth-death process for $\lambda=\mu$ is a
martingale. More meaningful information is obtained from the full
population distribution.  For simplicity in the ensuing formulas, we
now set $\lambda=\mu=1$ without loss of generality.  Let $P_n(t)$
denote the probability that the population consists of $n$ particles
at time $t$.  This probability distribution changes in time according
to
\begin{align}
  \label{P-dot}
  \dot P_n = (n-1)P_{n-1}-2n P_n+(n+1)P_{n+1}\,,
\end{align}
where we define $P_{-1}=0$, so that this equation is valid for all $n\geq 0$.
For the standard continuous-time random walk, the corresponding master
equation is $\dot P_n = P_{n-1}-2 P_n+P_{n+1}$.  We know that this random
walk eventually hits the origin, but that the average time to do so is
infinite.  We want to find the behavior of these two first-passage properties
for the birth-death process.

A convenient and powerful way to solve the master equation \eqref{P-dot} is
by the generating function method.  We first define the generating function
\begin{equation*}
  g(z,t)= \sum_{n\geq 0} P_n(t) z^n\,,
\end{equation*}
then take each of the equations for $\dot P_n$, multiply it by $z^n$, and
then sum over all $n$.  In doing so, we will encounter terms from the
right-hand side of \eqref{P-dot}, for example, the second term on the right,
that looks like
\begin{align*}
  \sum_{n\geq 0} 2n P_n\, z^n,
\end{align*}
which we can recast as
\begin{align*}
  2z\frac{\partial}{\partial z}  \sum_{n\geq 0}
  P_n\, z^n = 2z\frac{\partial g}{\partial z}\,.
\end{align*}
By this device of converting multiplication by $n$ to differentiation for all
three terms on the right-hand side of \eqref{P-dot}, we recast \eqref{P-dot}
as
\begin{equation}
  \label{gt}
  g_t = (z^2-2z+1)g_z= (1-z)^2g_z\,,
\end{equation}
where the subscripts now denote partial differentiation and the arguments of
$g$ are not written for compactness. 

This first-order partial differential equation can be simplified further by
defining the variable $y$ via $dy = dz/(1-z)^2$, which implies that
$y=1/(1-z)$, or equivalently, $z=1-y^{-1}$.  In terms of the variable $y$,
\eqref{gt} is converted to the classic wave equation $g_t=g_y$.  This
equation has the general solution $g=F(t+y)$, where the function $F$ is, in
principle, arbitrary, and whose explicit form is fixed by the initial
condition.  Let us specialize to the simple case of the single-particle
initial condition, namely, $P_n(t=0)=\delta_{n,1}$.  This immediately leads
to $g(z,t\!=\!0)=z$.  Then at $t=0$ the function $F$ is simply given by
$F(y)=z$.  However, we must express the right-hand side in terms of the true
dependent variable $y$, which means that $F(y)= 1-y^{-1}$.  Thus for any
$t\geq 0$ the generating function is
\begin{align}
  g(z,t)=F(y+t)= 1-\frac{1}{t+y}\,.
\end{align}

To extract the individual terms in the power-series representation of the
generating function, we now need to re-express this function in terms of $z$:
\begin{align}
  \label{g-series}
  g(z,t) &= 1 -\frac{1}{t+1/(1-z)} = 1- \frac{1-z}{(1+t)(1-zx)}\nonumber\\
         &= 1 -\frac{1-z}{1+t}\sum_{n\geq 0} (xz)^n\nonumber\\
  &=1- \frac{1}{1+t}\sum_{n\geq 0} (xz)^n+ \frac{z}{1+t}\sum_{n\geq 0}
    (xz)^n\,,
\end{align}
where, for notational simplicity, we introduce $x\equiv t/(1+t)$.  From the
last line of the above, we can immediately extract all the $P_n(t)$ and
obtain the well-known formulas:
\begin{align}
  \label{P-0n}
  P_0(t)= \frac{t}{1+t}\qquad\quad  P_n(t) =\frac{t^{n-1}}{(1+t)^{n+1}}\qquad n\geq 1\,.
\end{align}

With these results, we now obtain the first-passage properties of the
birth-death process.  The quantity $P_0(t)$ may be interpreted as the
probability that the population has gone extinct by time $t$, while
$S(t)=1-P_0(t)=1/(1+t)$ is the probability that the population survives up to
to time $t$.  Thus extinction is sure to occur, but the average extinction
time is infinite, just as for isotropic diffusion.  The main distinction with
isotropic diffusion is that $S(t)\sim t^{-1/2}$ for diffusion, while
$S(t)\sim t^{-1}$ for the birth-death process.  Thus survival is less likely
when the hopping rate is a linearly increasing function of $n$.

\section*{Concluding Comments}

These lecture notes have given a whirlwind tour through some basic and some
not-so-basic aspects of first-passage processes.  At the level of
fundamentals, I presented some classic results about first passage in the
simplest geometries of the infinite half line and the finite interval,
including first-passage probabilities, first-passage times, and splitting
probabilities.  I also discussed the intriguing connection between first
passage and electrostatics.  I then presented a number of applications.
Some, like the reaction rate of a cell and the birth-death process are
classic and have many immediate applications. Some, like the survival of a
diffusing lamb that is hunted by $N$ diffusing lions and survival of a
diffusing particle in a growing interval may seem somewhat idiosyncratic.
However, the solution methods are quite generic and may prove useful in many
other settings.  I hope that the uninitiated reader will enjoy learning about
some of these applications of first-passage processes and will be inspired to
delve further into this fascinating topic.

Much of the material in Secs.~\ref{sec:hunting} and
\ref{sec:growing-interval} stems from joint work with Paul Krapivsky.  I
thank him for pleasant collaborations on these projects, as well as pointing
out Ref.~\cite{E42} to me.  I also thank the National Science Foundation for
financial support over many years that helped advance some of the topics
discussed in these notes, most recently through NSF grant DMR-1910736.

\end{document}